\documentclass{emulateapj}

\usepackage{CJK}
\usepackage{natbib}
\usepackage{amsmath}
\usepackage{gensymb}
\usepackage{chngcntr}
\usepackage{longtable}

\shorttitle{Star formation laws in both Galactic massive clumps and external galaxies } \shortauthors{Liu et al.}

\begin{document}

\title{Star formation laws in both Galactic massive clumps and external galaxies: An extensive study with dust continuum, HCN (4-3), and CS (7-6)}
\author{Tie Liu \altaffilmark{1}, Kee-Tae Kim\altaffilmark{1}, Hyunju Yoo\altaffilmark{1}, Sheng-yuan Liu\altaffilmark{2}, Ken'ichi Tatematsu\altaffilmark{3}, Sheng-Li Qin\altaffilmark{4}, Qizhou Zhang\altaffilmark{5}, Yuefang Wu\altaffilmark{6}, Ke Wang\altaffilmark{7}, Paul F. Goldsmith\altaffilmark{8}, Mika Juvela\altaffilmark{9}, Jeong-Eun Lee\altaffilmark{10}, L. Viktor T\'{o}th\altaffilmark{11}, Diego Mardones\altaffilmark{12}, Guido Garay\altaffilmark{12},  Bronfman, Leonardo\altaffilmark{12}, Maria R. Cunningham\altaffilmark{13}, Di Li\altaffilmark{14}, Lo, Nadia\altaffilmark{12}, Isabelle Ristorcelli\altaffilmark{15}, Scott Schnee\altaffilmark{16} }

\altaffiltext{1}{Korea Astronomy and Space Science Institute 776, Daedeokdae-ro, Yuseong-gu, Daejeon, Republic of Korea 305-348; liutiepku@gmail.com}
\altaffiltext{2}{Academia Sinica, Institute of Astronomy and Astrophysics, P.O. Box 23-141, Taipei 106, Taiwan}
\altaffiltext{3}{National Astronomical Observatory of Japan, 2-21-1 Osawa, Mitaka, Tokyo 181-8588, Japan}
\altaffiltext{4}{Department of Astronomy, Yunnan University, and Key Laboratory of Astroparticle Physics of Yunnan Province, Kunming, 650091, China}
\altaffiltext{5}{Harvard-Smithsonian Center for Astrophysics, 60 Garden Street, Cambridge, MA 02138, USA}
\altaffiltext{6}{Department of Astronomy, Peking University, 100871, Beijing China}
\altaffiltext{7}{European Southern Observatory, Karl-Schwarzschild-Str.2, D-85748 Garching bei M\"{u}nchen, Germany}
\altaffiltext{8}{Jet Propulsion Laboratory, California Institute of Technology, 4800 Oak Grove Drive, Pasadena, CA 91109, USA 0000-0002-6622-8396}
\altaffiltext{9}{Department of physics, University of Helsinki, FI-00014 Helsinki, Finland}
\altaffiltext{10}{School of Space Research, Kyung Hee University, Yongin-Si, Gyeonggi-Do 446-701, Korea}
\altaffiltext{11}{Department of Astronomy of the Lor\'{a}nd E\"{o}tv\"{o}s University, P\'{a}zm\'{a}ny P\'{e}ter s\'{e}t\'{a}ny 1, 1117 Budapest, Hungary}
\altaffiltext{12}{Departamento de Astronom\'{\i}a, Universidad de Chile, Casilla 36-D, Santiago, Chile}
\altaffiltext{13}{School of Physics, University of New South Wales, Sydney, NSW 2052, Australia}
\altaffiltext{14}{National Astronomical Observatories, Chinese Academy of Sciences, Beijing, 100012, China}
\altaffiltext{15}{IRAP, CNRS (UMR5277), Universit\'{e} Paul Sabatier, 9 avenue du Colonel Roche, BP 44346, 31028, Toulouse Cedex 4, France}
\altaffiltext{16}{National Radio Astronomy Observatory, 520 Edgemont Road, Charlottesville, VA 22903}

\begin{abstract}
We observed 146 Galactic clumps in HCN (4-3) and CS (7-6) with the Atacama Submillimeter Telescope Experiment (ASTE) 10-m telescope.  A tight linear relationship between star formation rate and gas mass traced by dust continuum emission was found for both Galactic clumps and the high redshift ($\textit{z}>1$) star forming galaxies (SFGs), indicating a constant gas depletion time of $\sim$100 Myr for molecular gas in both Galactic clumps and high $\textit{z}$ SFGs. However, low $\textit{z}$ galaxies do not follow this relation and seem to have a longer global gas depletion time. The correlations between total infrared luminosities (L$_{TIR}$) and molecular line luminosities (L$'_{mol}$) of HCN (4-3) and CS (7-6) are tight and
sublinear extending down to clumps with L$_{TIR}\sim10^{3}$ L$_{\sun}$. These correlations become linear when
extended to external galaxies. A bimodal behavior in the L$_{TIR}$--L$'_{mol}$ correlations was found for clumps
with different dust temperature, luminosity-to-mass ratio, and $\sigma_{line}/\sigma_{vir}$. Such bimodal behavior may
be due to evolutionary effects. The slopes of L$_{TIR}$--L$\arcmin_{mol}$ correlations become more shallow as clumps evolve. We compared our results with lower J transition lines in \cite{wu10}. The correlations between clump
masses and line luminosities are close to linear for low effective excitation density tracers but become sublinear
for high effective excitation density tracers for clumps with L$_{TIR}$ larger than L$_{TIR}\sim10^{4.5}$ L$_{\sun}$. High effective excitation density tracers cannot linearly trace the total clump masses, leading to a sublinear correlations for both M$_{clump}$--L$\arcmin_{mol}$ and L$_{TIR}$–-L$\arcmin_{mol}$ relations.
\end{abstract}


\section{Introduction}

The empirical correlation between the star formation rate (SFR) surface
density ($\Sigma_{SFR}$) and the surface density of cold gas ($\Sigma_{gas}$) pioneered in the works of \cite{sch59} \& \cite{ken98}, the so-called Kennicutt--Schmidt (K-S) law, is of
great importance for input into theoretical models of galaxy evolution. The power-law relations between $\Sigma_{SFR}$ and total gas surface density ($\Sigma_{gas}=\Sigma_{HI}+\Sigma_{H_{2}}$) used to
describe star formation across entire galaxies and galactic nuclei, have a
typical power-law index of $\sim$1.4--1.6 \citep{ken98,ken12}.

However, since stars form in dense cores of molecular clouds, the relations between $\Sigma_{SFR}$ and total molecular gas surface density ($\Sigma_{mol}$) may be more fundamental to describe star formation. Indeed, very tight relations between SFR indicated by infrared luminosities (L$_{IR}$) and dense molecular gas mass indicated by molecular line luminosities (L$\arcmin_{mol}$) have been revealed in external galaxies \citep{gao04,gre14,zhang14,liu15}.

Interestingly, the L$_{IR}$--L$\arcmin_{mol}$ relations are close to linear for J = 1--0 up to J = 5--4 of CO \citep{gre14}. From J = 6--5 and up to the J = 13--12 transition of CO, \cite{gre14} found an increasingly sublinear slope with increasing J. They argued that the thermal state of high J CO transitions is unlikely to be maintained by star-formation-powered far--UV radiation fields and thus is no longer directly tied to the star formation rate. Such trend is similar to the predictions in simulations \citep{nara08}.

However, from Herschel SPIRE FTS observations of 167 local
galaxies, \cite{liu15} found that the L$_{IR}$--L$\arcmin_{mol}$ relations are essentially linear and tight for all the CO transitions including the highest J=12-11 transition, indicating that the SFR is linearly correlated with the dense molecular gas. \cite{liu15} argued that the non-linear result reported in \cite{gre14} can be attributed to the comparatively small number of galaxies in their sample.

Similarly, nearly linear L$_{IR}$--L$\arcmin_{mol}$ correlations for dense molecular gas tracers (e.g. HCN and CS) were found toward both Galactic dense clumps and galaxies \citep{wu05,wu10,ma13,zhang14,gao04}, indicating a constant SFR per unit mass from the scale of dense clumps to that of distant galaxies \citep{wu05,wu10}. And dense molecular line luminosity seem to be linearly related to the mass of dense gas most relevant to star
formation \citep{wu10,reiter11}. Through observations of nearby clouds, \cite{lada10} found that SFR in molecular clouds seems to be linearly proportional to the cloud mass above an extinction threshold of A$_{K}\approx$0.8 mag. If star formation only depends on the mass of dense gas, the underlying star formation scaling law will be always linear for clouds and galaxies with the same dense gas fraction \citep{lada12}.

However, apparently sublinear L$_{IR}$--L$\arcmin_{mol}$ correlations were found for middle J dense line tracers like HCN (3-2) and CS (7-6) in Galactic clumps \citep{wu10}. In contrast, linear L$_{IR}$--L$\arcmin_{mol}$ correlations for J=7-6 of CS, J=4-3 of HCN and HCO$^{+}$ were found in nearby star-forming galaxies \citep{zhang14}. Are the linear observed Kennicutt--Schmidt scaling relations for dense gas in galaxies an artifact of unresolved measurements of GMCs due to the beam dilution as suggested by \cite{lada13}?

In spite of dense molecular gas tracers, optically thin dust continuum emission at (sub)millimeter wavelengths has often been used as a tracer for molecular gas masses in extragalactic studies \citep{dunne00,scov16}. \cite{dunne00} compared the dust masses, derived from the submillimetre fluxes, to the H$_{2}$, H{\sc i} and H$_{2}$+H{\sc i} masses in 104 Local Universe galaxies. The strongest correlation is with the molecular gas, which suggests that the dust is primarily found in molecular clouds \citep{dunne00}. Strong correlations between dust masses and star formation rates have been revealed in both low $\textit{z}$ and high $\textit{z}$ galaxies \citep{dunne00,da10,row12,clem10,clem13,mag13,sant14,hjor14,scov16}. In this work, we will also explore the star formation law in Galactic massive clumps with dust continuum emission and compare with extragalactic studies.

\section{Observations}

\subsection{The sample}

The sample of 146 sources in the survey is obtained from \cite{fa04}. The basic parameters (like coordinates, sizes, masses, temperature, bolometric luminosities, densities) of these sources were summarized in Table 1 of \cite{fa04}. The 146 sources are IRAS point sources and the main clumps in 1.2 mm continuum emission, which have far infrared colors typical of Ultra-Compact H{\sc ii} (UC H{\sc ii}) regions. Most of the sources (91\%) have bolometric luminosities $\geq10^{4}$ L$_{\sun}$, indicating that they harbor at least one B0.5 type massive star \citep{fa04}. The median and maximum bolometric luminosities of this sample are $6.4\times10^{4}$ and 5.6$\times10^{6}$ L$_{\sun}$, respectively. The dust temperature ranges from 18 to 46 K with a median value of 31 K \citep{fa04}. The volume densities of these sources, which were derived assuming that the clumps have spherical morphologies, range from 5.5$\times10^{3}$ to $3.8\times10^{6}$ with a median value of $\sim1\times10^{5}$ cm$^{-3}$, spanning three orders of magnitude \citep{fa04}. Considering the large diversity of both densities and bolometric luminosities, this sample is ideal for studies of the Kennicutt--Schmidt law in the Milky Way. \\

\subsection{Observations with the ASTE 10 m telescope}

The single pointing observations toward the 146 Galactic clumps were conducted with the Atacama Submillimeter Telescope Experiment (ASTE) 10-m telescope with position switching mode between July 1st and July 4th in 2014. The pointing positions were presented in Table 1 of \cite{fa04}. The two-sideband single-polarization heterodyne receiver DASH345/CATS345, operating at frequencies of 324-372 GHz was used to observe HCN (4-3) and CS (7-6) lines simultaneously. We used the MAC spectrometer with a spectral resolution of 0.5 MHz or 0.42 km~s$^{-1}$. The beam size at 354 GHz is $\sim22\arcsec$. The main beam efficiency was $\sim$0.6. The system temperature varied from 400 to 500 K during observations. The rms level per channel is $\sim$0.1-0.2 K in antenna temperature. The data were reduced with Gildas/Class package \citep{gui00}. Baseline subtraction was performed by fitting linear functions.\\

\section{Results}

\subsection{Observational results of Molecular lines}

We detected HCN (4-3) toward 141 clumps and CS (7-6) toward 137 clumps. In Figure 1, we show the HCN (4-3) and CS (7-6) spectra of two sources for example. For IRAS 18317-0757, both HCN (4-3) and CS (7-6) spectra can be well fitted with single Gaussian profiles. For IRAS 09018-4816, CS (7-6) has a Gaussian profile, while HCN (4-3) shows asymmetric profiles with an absorption dip. There are 74 ($\sim$52\%) clumps showing asymmetric profiles in HCN (4-3) like IRAS 09018-4816. While there are only 22 (16\%) clumps showing asymmetric profiles in CS (7-6). The line profiles were classified into three categories as shown in the 6th and 11th columns in Table 1. Those denoted with ``G" have symmetric profiles which can be well fitted with Gaussian profiles. Those denoted with ``B" or ``R" have asymmetric profiles with their observed emission peaks blueshifted or redshifted with respect to the peak velocities derived from Gaussian fits, respectively. In this work, we will not discuss the line profiles in details. The line profiles will be modelled and discussed in another paper for kinematics (e.g., infall and outflow) studies.

Many lines especially HCN (4-3) lines with asymmetric profiles show absorption dips due to self-absorption. Previous works like \cite{wu05,wu10} and \cite{step16} simply used Moment 0 method to calculating the integrated intensity of such optically thick line, which may severely underestimate the total emission including the blocked radiation due to self-absorption \citep{van09,Oss10}. Gaussian fits were usually performed to obtain a rough correction for self-absorption and \textbf{to} compare with radiation transfer models \citep{van09,Oss10}. In this work, we fitted all the detected spectra with Gaussian functions and present the results (integrated intensity $\int T_{A}^{*}dV$, systemic velocity V$_{lsr}$, the full width of half maximum (FWHM) and peak antenna temperature T$_{A}^{*}$) in Table 1. Due to self-absorption, most HCN (4-3) lines have larger errors in the integrated intensities from Gaussian fit than CS (7-6) lines. In Figure 2, we compare the integrated intensities inferred from Gaussian fits with the values simply obtained from integrating the lines (Moment 0). The integrated intensities obtained from different methods are very consistent and roughly linearly correlated. Therefore, we claim that the integrated intensities obtained from Gaussian fits are reliable.

The median line widths for HCN (4-3) lines with symmetric profiles and with asymmetric profiles are $\sim$5.3 and $\sim$8.1 km~s$^{-1}$, respectively. The median line widths for CS (7-6) lines with symmetric profiles and with asymmetric profiles are $\sim$4.0 and $\sim$7.5 km~s$^{-1}$, respectively. In general, lines with asymmetric profiles have much larger line widths than lines with symmetric profiles, indicating that those clumps with asymmetric line profiles may be more turbulent and more likely be affected by stellar feedback or bulk motions (e.g. infall, outflows). CS (7-6) lines usually have smaller line widths than HCN (4-3). The median line width ratio of CS (7-6) to HCN (4-3) is $\sim$0.67.\\

\begin{figure*}
\centering
\includegraphics[angle=90,scale=0.5]{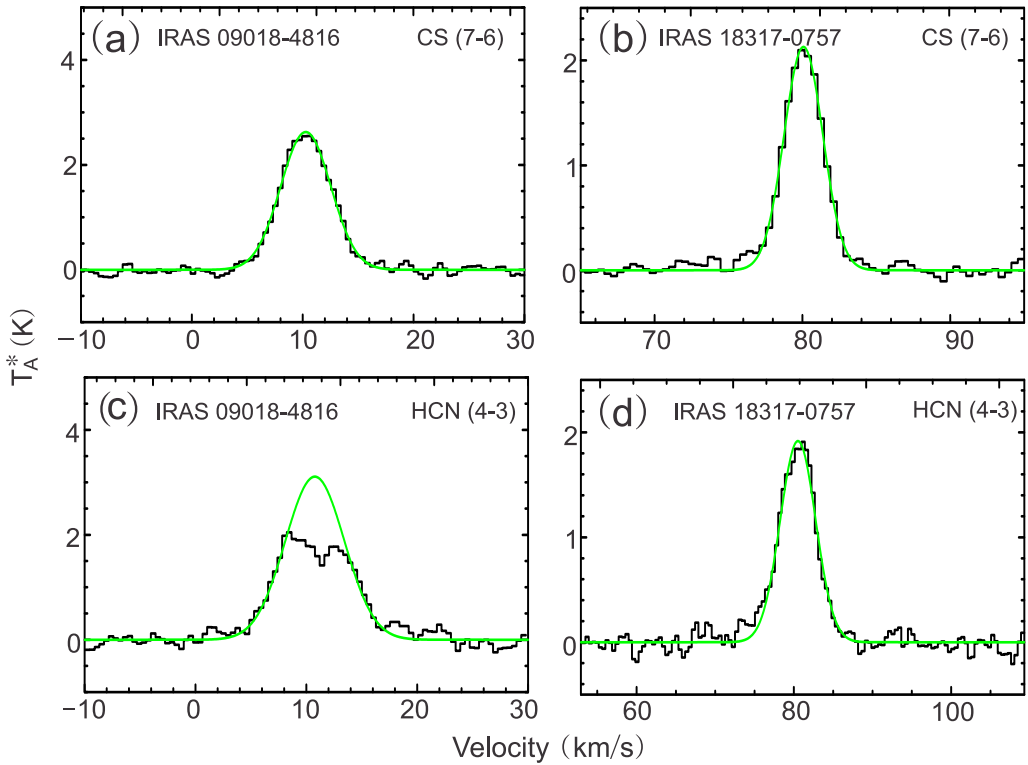}
\caption{Spectra of IRAS 09018-4816 (left panels) and IRAS 18317-0757 (right panels). The upper panels show CS (7-6) lines. The lower panels show HCN (4-3) lines. The green lines are Gaussian fits.}
\end{figure*}

\begin{figure}
\centering
\includegraphics[angle=0,scale=0.48]{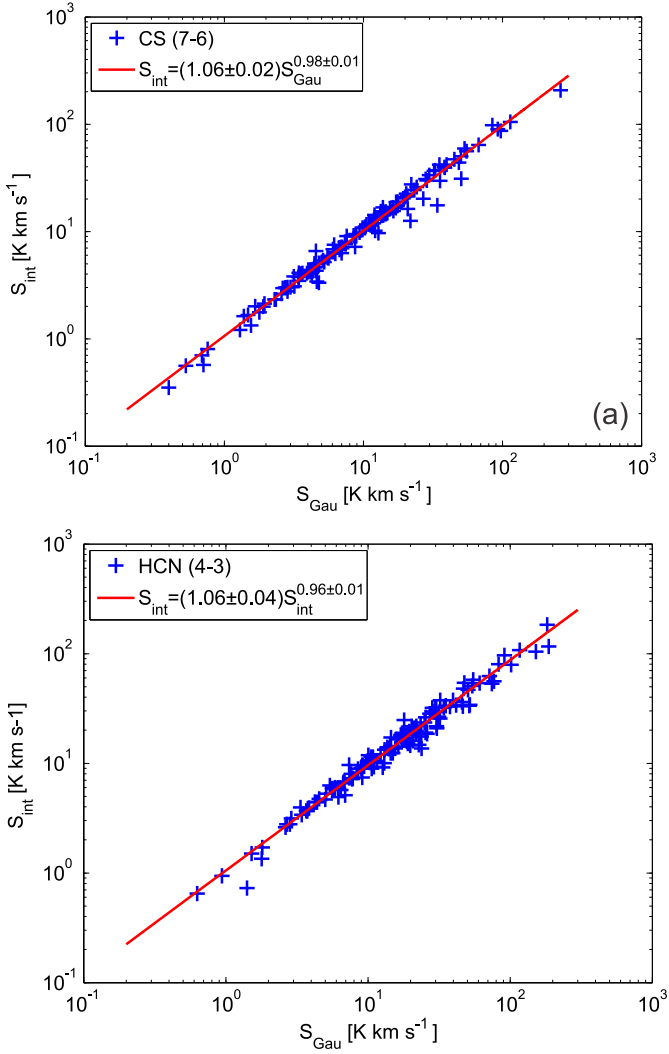}
\caption{Comparison between the integrated intensities (S$_{Gau}$) from Gaussian fits and the values (S$_{int}$) simply from integrating the lines (Moment 0). The red lines represent the best power-law fits to the data. Panel (a): CS (7-6). Panel (b): HCN (4-3).}
\end{figure}

\subsection{Line luminosities}

Assuming a Gaussian brightness distribution for the
source and a Gaussian beam, molecular line luminosities $L\arcmin_{mol}$ can be derived following \cite{wu10}
\begin{equation}
\begin{split}
L\arcmin_{mol}=23.5\times10^{-6}\times D^{2}\times (\frac{\pi\times\theta_{s}^{2}}
{4ln 2})\times(\frac{\theta_{s}^{2}+\theta_{beam}^{2}}{\theta_{s}^{2}}) \\
\times \int T^{*}_{A}dV/\eta
\end{split}
\end{equation}
Here D is the distance in kpc obtained from \cite{fa04}, and $\theta_{s}$ and $\theta_{beam}$
are the size of the line emission source and of the beam in
arcsecond, and $\eta$ is the main beam efficiency. Since we only carried out single pointing observations and thus can not determine the exact line emission area, we here assume that the source sizes of HCN (4-3) and CS (7-6) are the same as that of 1.2 mm continuum emission in \cite{fa04}. The caveats in estimating line luminosity will be discussed in section 4.1.

The derived HCN (4-3) ($L'_{HCN}$) and CS (7-6) ($L'_{CS}$) luminosities are listed in the 4th and 5th columns of Table 2. The median values of $L'_{HCN}$ and $L'_{CS}$ are $\sim$16.4 and $\sim$9.2 K~km~s$^{-1}$~pc$^{2}$, respectively. The mean values of $L'_{HCN}$ and $L'_{CS}$ are $\sim$62.5 and $\sim$56.6 K~km~s$^{-1}$~pc$^{2}$, respectively.

\subsection{Gravitational Stability of dense clumps}

To examine the gravitational stability of those dense clumps, we derive one dimensional
virial velocity dispersions ($\sigma_{vir}$):
\begin{equation}
\sigma_{vir}=\sqrt{\frac{\gamma M_{clump}G}{5R_{clump}}}
\end{equation}
here M$_{clump}$ is the clump mass, G is the gravitational constant, and R$_{clump}$ is the radius of the dense clump. M$_{clump}$ and R$_{clump}$ were taken from \cite{fa04}. Parameter $\gamma$ is the geometric factor equal to unity for a
uniform density profile and 5/3 for an inverse square profile \citep{wil94}. Here we take $\gamma$ equal to 5/3. The derived one dimensional
virial velocity dispersions ($\sigma_{vir}$) are listed in the 6th column of Table 2.

Assuming the line profiles are Gaussian, the one dimensional velocity dispersions of molecular lines are
\begin{equation}
\sigma_{line}=\frac{FWHM}{2\sqrt{2ln(2)}}=\frac{FWHM}{2.355}.
\end{equation}
We derived $\sigma_{HCN}$ and $\sigma_{CS}$ for HCN (4-3) and CS (7-6), respectively. The dense clumps may be under gravitational collapse if $\sigma_{vir}$ is larger than $\sigma_{line}$. The median ratio of $\sigma_{HCN}$ to $\sigma_{vir}$ is $\sim$1.2, and the median ratio of $\sigma_{CS}$ to $\sigma_{vir}$ is $\sim$0.8, indicating that most sources should be in virial equilibrium. There are 45 dense clumps having both $\sigma_{HCN}$ and $\sigma_{CS}$ smaller than $\sigma_{vir}$, indicating that the gas motions may not provide sufficient support against the local gravitational collapse in these clumps. In contrast, there are 37 dense clumps that have both $\sigma_{HCN}$ and $\sigma_{CS}$ larger than $\sigma_{vir}$, indicating that these clumps may be more turbulent and unbound. However, a broad line profile can also be caused by infall. Among the 37 sources, 11 show ``blue profile" with blueshifted emission peak stronger than redshifted one, a signature for collapse \citep{zhou93}, indicating that these 11 sources may be also bound and probably be in global collapse. Further mapping studies will tell whether they are really in collapse or not.  \\

\subsection{The Kennicutt-Schmidt law revealed by 1.2 mm continuum}

\begin{figure*}
\centering
\includegraphics[angle=90,scale=0.6]{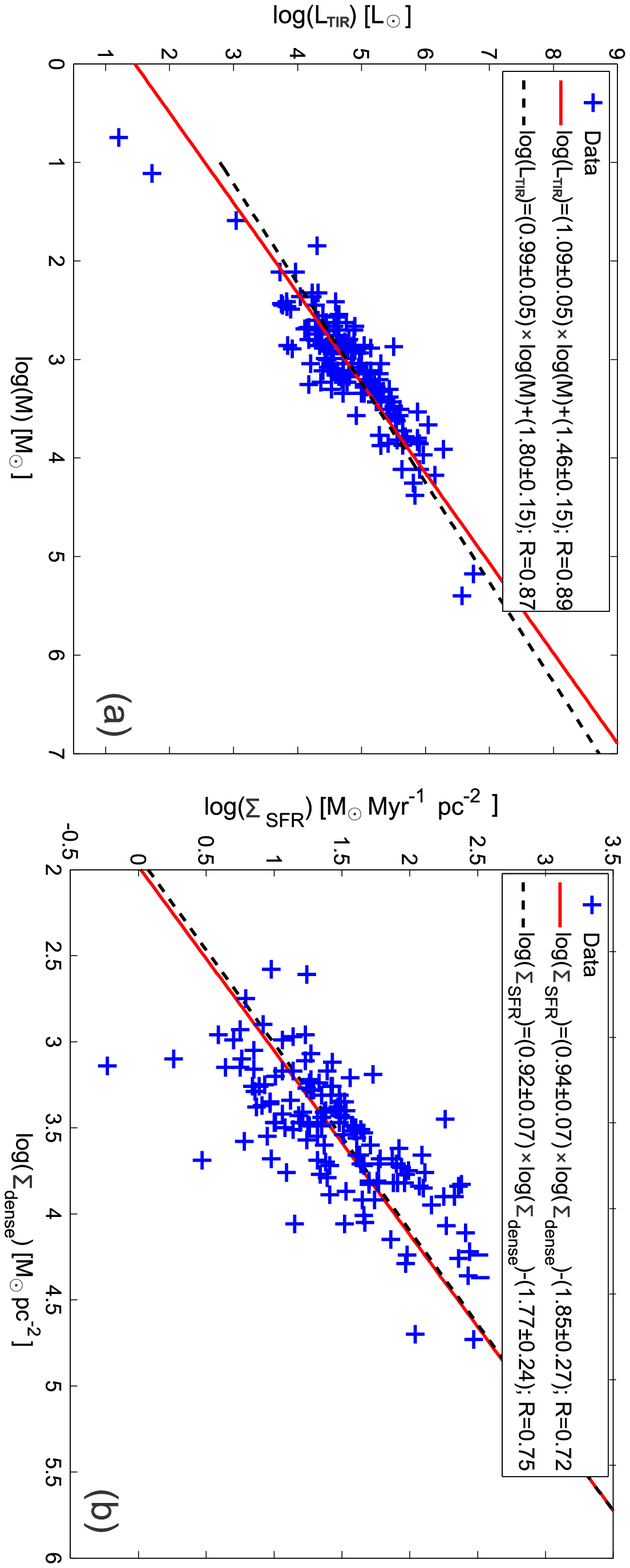}
\caption{(a). Infrared luminosities L$_{TIR}$ vs. Clump masses M (for simplicity, hereafter we use M rather than M$_{clump}$ in the figures). The red line represents least squares fit to all the sources. The black dashed line represents least squares fit to the clumps with L$_{TIR}$ larger than 10$^{3}$ L$_{\sun}$. (b). Surface density of star formation rate, $\Sigma_{SFR}$, against surface density of the dense molecular gas mass as traced by the 1.2 mm continuum,
$\Sigma_{dense}$. The red line represents least squares fit to all the sources. The black dashed line represents least squares fit to the clumps with L$_{TIR}$ larger than 10$^{3}$ L$_{\sun}$. }
\end{figure*}

\begin{figure*}
\centering
\includegraphics[angle=-90,scale=0.6]{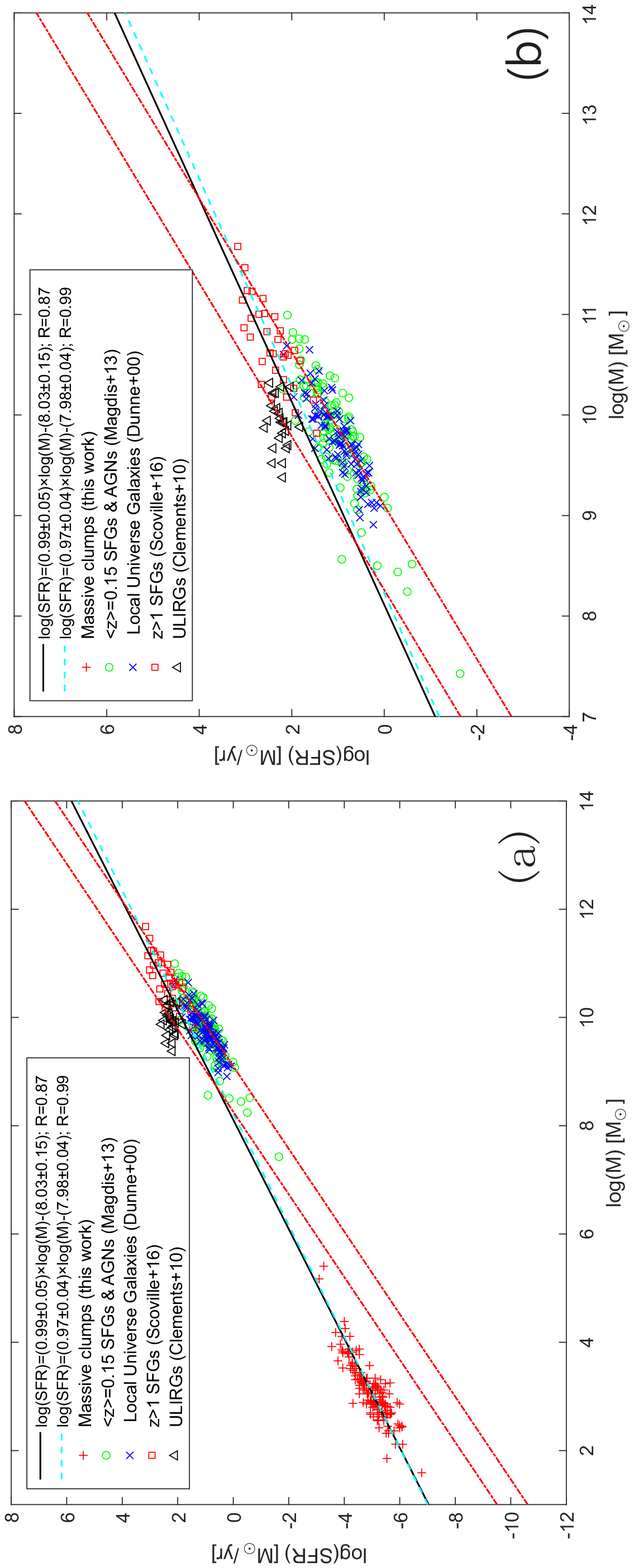}
\caption{(a). SFR vs. gas mass correlations. Red crosses are Galactic massive clumps from this work; Green open circles are star-forming galaxies (SFGs) and active galactic nuclei (AGNs) with intermediate redshift ($<\textit{z}>$=0.15) from \cite{mag13}; blue $\times$ signs are Local Universe Galaxies from \cite{dunne00}; red open squares are high redshift ($\textit{z}>$1) SFGs from \cite{scov16} and black triangles are local ultraluminous infrared galaxies (ULIRGs) from \cite{clem10}. The black line represents power law fit toward the Galactic massive clumps only. The cyan dashed line represents power fit toward both Galactic massive clumps and the stacked SFGs samples in \cite{scov16}. The SFR in all works are re-calibrated with equation (5) in section 3.4. We convert the dust mass to gas mass with a constant dust-to-gas ratio of 0.01 \citep{sant14}. The two red dashed lines correspond to the integrated Schmidt-Kennicutt laws fitted by \cite{daddi10} for both local spirals and distant BzK galaxies (lower line), and for ULIRGs (upper line). (b). SFR vs. gas mass correlations for external galaxies only. The marks and lines are the same as in panel (a). }
\end{figure*}

In panel (a) of Figure 3, we first examine the correlation between total infrared luminosities (L$_{TIR}$) and Clump masses ($M_{clump}$) for Galactic clumps. L$_{TIR}$ was obtained from fit of the SED (consisting
of the 1.2 mm flux and the fluxes in the four IRAS bands), and $M_{clump}$ was derived from 1.2 mm continuum \citep{fa04}. In section 4.1, we will show that the L$_{TIR}$ obtained from integrating the SED in our work are consistent with the total infrared luminosities derived only from the four IRAS bands as used in other works\citep{wu05,wu10}. We use L$_{IR}$ for the total infrared luminosities derived only from the four IRAS bands to distinguish from the total infrared luminosities derived from SED fit (L$_{TIR}$). The least squares fit to all the data gave a super-linear slope due to the contamination of several low luminosity sources which are located obviously below the fit. The clumps with luminosities lower than 10$^{3}$ L$_{\sun}$ are likely low-mass star forming regions. If we only consider the clumps with L$_{TIR}$ larger than 10$^{3}$ L$_{\sun}$, the least squares fit indicates that the relation between L$_{TIR}$ and $M$ is linear. The linear relation between L$_{TIR}$ and $M_{clump}$ can be expressed as:
\begin{equation}
\frac{L_{TIR}}{L_{\sun}}=10^{1.80\pm0.15}\frac{M_{clump}}{M_{\sun}}
\end{equation}
Following \cite{ken12}, we can convert L$_{TIR}$ to a star formation rate (SFR; $\dot{M}_{*}$) as:
\begin{equation}
\textrm{log}(\frac{\dot{M}_{*}}{M_{\sun}~yr^{-1}})=\textrm{log}(\frac{L_{TIR}}{ergs~s^{-1}})-43.41
\end{equation}
The use of extragalactic relation between L$_{TIR}$ and SFR may significantly underestimate the SFR for low-mass molecular clumps, where the initial mass function (IMF) is not fully sampled \citep{wu05,Vuti13}.
While L$_{TIR}$ can well trace SFR for dense clumps having L$_{TIR}$ exceed 10$^{4.5}$ L$_{\sun}$ \citep{wu05,Vuti13}. In our sample, 71\% sources have L$_{TIR}$ exceed 10$^{4.5}$ L$_{\sun}$ and 92\% have L$_{TIR}$ exceed 10$^{4}$ L$_{\sun}$. In addition, we do not see clear breakup in the L$_{TIR}$ and $M_{clump}$ correlation. Therefore, the applying of extragalactic relation on our sample may not severely underestimate the SFR.

From Equation (4) and (5), we can derive the relation between $\dot{M}_{*}$ and $M$:
\begin{equation}
\frac{\dot{M}_{*}}{M_{\sun}~yr^{-1}}=(0.93^{+0.38}_{-0.27})\times10^{-8}\frac{M_{clump}}{M_{\sun}}
\end{equation}
Then, the gas depletion time $\tau_{dep}$ is:
\begin{equation}
\tau_{dep}=\frac{M_{clump}}{\dot{M}_{*}}=107^{+44}_{-31}~Myr
\end{equation}

\cite{wu05} derived a correlation between $\dot{M}_{*}$ and $M_{clump}$ as $\frac{\dot{M}_{*}}{M_{\sun}~yr^{-1}}\sim1.2\times10^{-8}\frac{M_{clump}}{M_{\sun}}$, corresponding to
a gas depletion time of $\sim$83 Myr, which is smaller than the value we derived here. The difference is because that they used a larger conversion factor ($\frac{\dot{M}_{*}}{M_{\sun}~yr^{-1}}=2.0\times10^{-10}\frac{L_{IR}}{L_{\sun}}$) between infrared luminosities and star formation rates. If we use the same conversion factor, the depletion time
should be $\sim79^{+33}_{-23}$ Myr, which is consistent with their value.

To test the Kennicutt-Schmidt law in terms of the corresponding surface densities of SFR ($\Sigma_{SFR}$) and dense gas ($\Sigma_{dense}$), we normalized SFR and M$_{clump}$
using the sizes of the 1.2 mm continuum emission from \cite{fa04} to obtain $\Sigma_{SFR}$ and $\Sigma_{dense}$. In panel (b) of Figure 3, we present the derived $\Sigma_{SFR}$ and $\Sigma_{dense}$ for the whole sample.
The correlation between $\Sigma_{SFR}$ and $\Sigma_{dense}$ for the whole sample is nearly linear. But the relation for the sources with L$_{TIR}$ larger than 10$^{3}$ L$_{\sun}$ seems to have sub-linear slope. In general, the $\Sigma_{SFR}$--$\Sigma_{dense}$ correlation is not as tight as the L$_{TIR}$--M$_{clump}$ correlation.

In Figure 4, we investigate the integrated Schmidt-Kennicutt laws (SFR vs. gas mass correlations) for both Galactic massive clumps and external galaxies. The external galaxies samples are star-forming galaxies (SFGs) and active galactic nuclei (AGNs) with intermediate redshift ($\textit{z}\sim$0.15) from \cite{mag13}, Local Universe Galaxies \citep{dunne00}, high redshift ($\textit{z}>$1) SFGs \citep{scov16} and local ultraluminous infrared galaxies (ULIRGs) from \cite{clem10}.
We convert the total infrared luminosities of these external galaxies to SFR with equation (5). Their dust masses, which were derived from (sub)millimeter continuum, were converted to gas masses by assuming a constant dust-to-gas ratio of 0.01 \citep{sant14}. The dependence of the gas metallicity with dust-to-gas ratio was not considered. This, however, may introduce only a minor effect on correlations because the gas metallicity changes less than a factor of 2-3, while the dust mass spans 2-3 orders of magnitude for these external galaxies \citep{sant14}. As shown by the black line, we extend the SFR vs. gas mass correlation of Galactic massive clumps to external galaxies. Interestingly, we find that all external galaxies except for high redshift ($\textit{z}>$1) SFGs significantly deviate from the correlation. The intermediate redshift star-forming galaxies (SFGs) and active galactic nuclei (AGNs) \citep{mag13}, and Local Universe Galaxies \citep{dunne00} are located below the correlation, indicating that they have smaller star forming efficiencies or longer gas depletion time than Galactic massive clumps. While ULIRGs, which have very high star forming efficiencies, are located above the correlation. Interestingly, high redshift ($\textit{z}>$1) SFGs seem to follow the correlation very well. We find a very tight linear correlation between SFR and gas mass for both Galactic massive clumps and the stacked high redshift SFGs. The linear relation can be expressed as:
\begin{equation}
\textrm{log}(\frac{SFR}{M_{\sun}/yr})=(0.97\pm0.04)\textrm{log}(\frac{M}{M_{\sun}})-(7.98\pm0.04); R=0.99
\end{equation}
This tight relation may indicate a constant molecular gas depletion time of $\sim$100 Myr for both Galactic massive clumps and high redshift SFGs.\\

\subsection{The dense molecular gas Kennicutt-Schmidt law}

\begin{figure*}
\centering
\includegraphics[angle=90,scale=0.7]{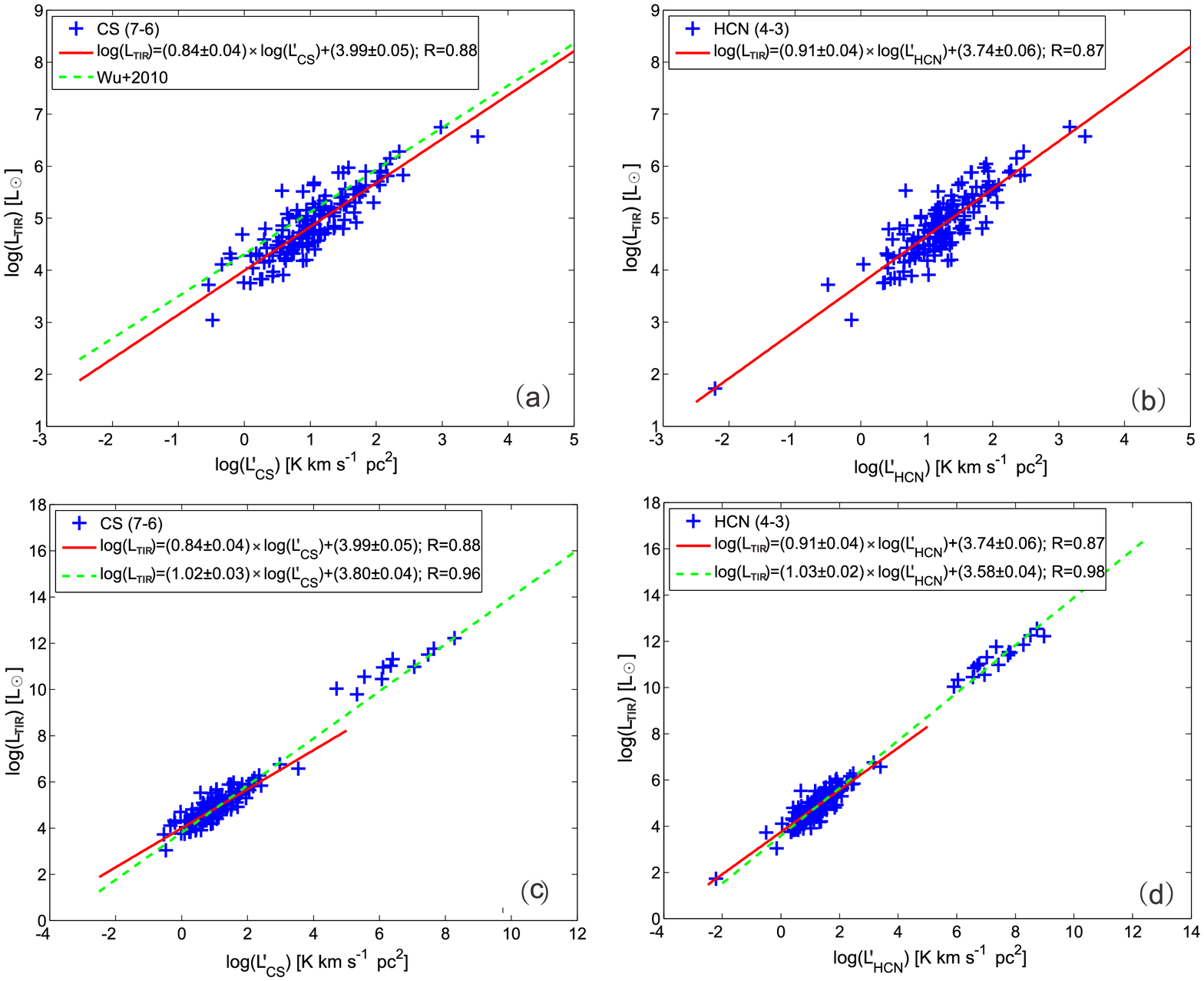}
\caption{Upper-left panel: L$_{TIR}$--L$\arcmin_{CS}$ correlation for Galactic clumps. Upper-right panel: L$_{TIR}$--L$\arcmin_{HCN}$ correlation for Galactic clumps. Lower-left panel: Upper-left panel: L$_{TIR}$--L$\arcmin_{CS}$ correlation for Galactic clumps and active galaxies. Lower-right panel: L$_{TIR}$--L$\arcmin_{HCN}$ correlation for Galactic clumps and active galaxies. The red lines in each panel represent least squares fits toward Galactic clumps. The green dashed line in the upper left panel represent the least squares fits toward Galactic clumps from Wu et al. (2010). The green dashed lines in the lower panels represent the least squares fits toward both Galactic clumps and active Galaxies. }
\end{figure*}

\begin{figure}
\centering
\includegraphics[angle=0,scale=0.47]{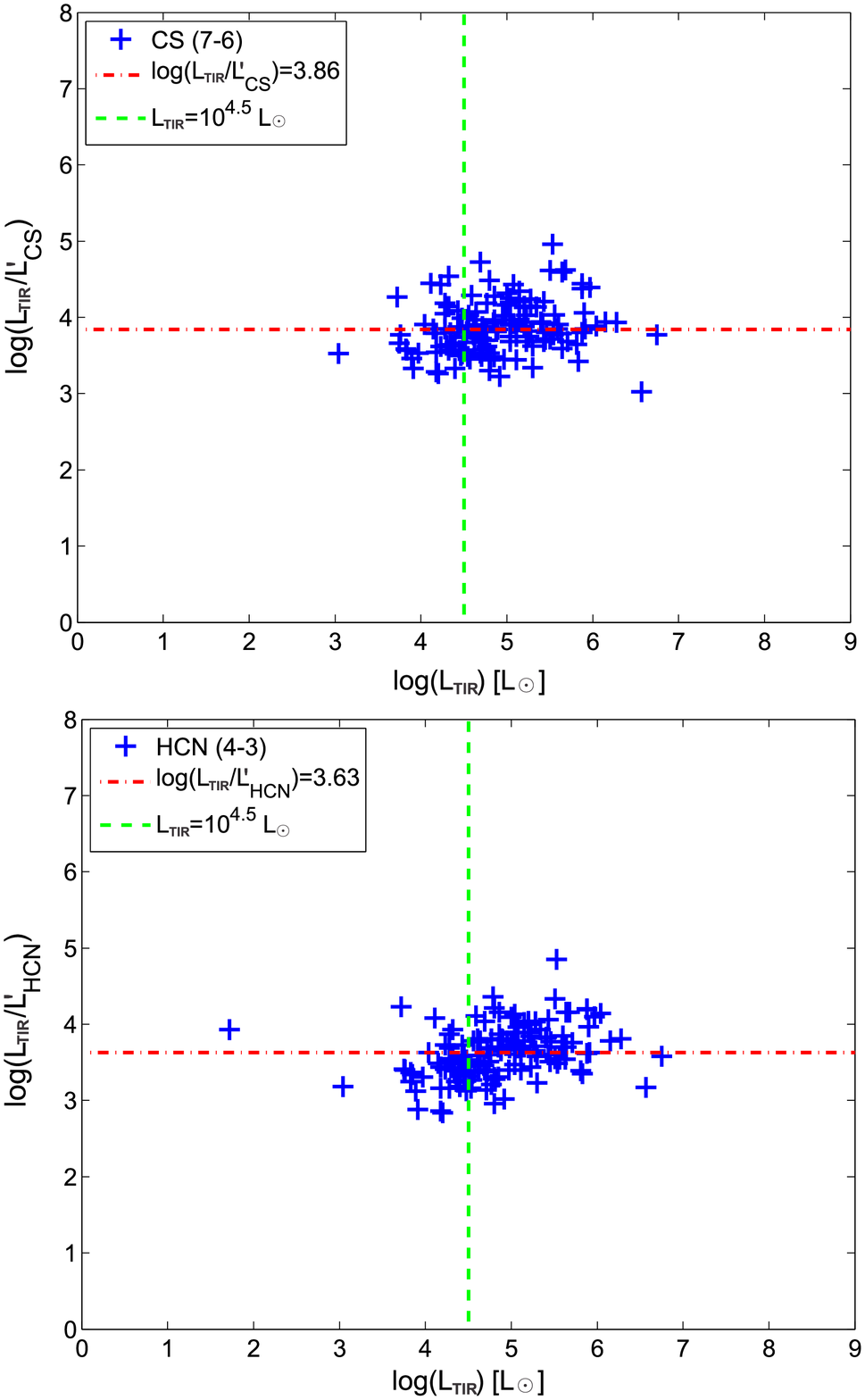}
\caption{Correlations between L$_{TIR}$/L$\arcmin_{mol}$ and
L$_{TIR}$ for CS (7-6) (Upper panel) and HCN (4-3) (Lower panel). The horizontal dashed
lines indicate the median L$_{TIR}$/L$\arcmin_{mol}$ ratio. The vertical dashed lines show the cutoff at L$_{TIR}=10^{4.5}$ L$_{\sun}$, suggested in \cite{wu05,wu10}. }
\end{figure}

In panels (a) and (b) of Figure 5, we show the L$_{TIR}$--L$\arcmin_{CS}$ and L$_{TIR}$--L$\arcmin_{HCN}$ correlations for Galactic clumps, respectively. Both L$_{TIR}$--L$\arcmin_{CS}$ and L$_{TIR}$--L$\arcmin_{HCN}$ correlations are very tight over about four orders of magnitude in L$_{TIR}$. The best linear least-squares fits with uncertainties are listed
below, with R indicating the correlation coefficient:
\begin{equation*}
\begin{split}
\textrm{log}(L_{TIR})=(0.84\pm0.04)\times \textrm{log}(L\arcmin_{CS})\\+(3.99\pm0.05);~R=0.88\\
\textrm{log}(L_{TIR})=(0.91\pm0.04)\times \textrm{log}(L\arcmin_{HCN})\\+(3.74\pm0.06);~R=0.87
\end{split}
\end{equation*}
From Bayesian regression with LINMIX\_ERR (Kelly 2007), the slopes of L$_{TIR}$--L$\arcmin_{CS}$ and L$_{TIR}$--L$\arcmin_{HCN}$ are 0.85$\pm$0.04 and 0.93$\pm$0.05, respectively, which are consistent with results from linear least-squares fits. The relations are sub-linear. \cite{wu10} found a slope of 0.81$\pm$0.04 for L$_{IR}$--L$\arcmin_{CS}$ relation, which is similar to the slope derived for our sample. However, as shown in green dashed line in panel (a), the intercept in their correlation is larger than ours. We will discuss this inconsistence in section 4.1.

A threshold in luminosity at L$_{IR}=10^{4.5}$L$_{\sun}$ was found for the correlations between L$_{IR}$ and molecular line luminosity from observations of lower J transition lines \citep{wu05,wu10}. Above the threshold, a nearly linear correlation was found between the infrared luminosity and the line luminosity of all dense gas tracers for Galactic dense clumps \citep{wu05,wu10}. In Figure 6, the distance independent ratio L$_{TIR}$/L$\arcmin_{mol}$ has been plotted versus
L$_{TIR}$. In general, sources with L$_{TIR}>10^{4.5}$L$_{\sun}$ have slightly larger L$_{TIR}$/L$\arcmin_{mol}$ than sources with L$_{TIR}<10^{4.5}$L$_{\sun}$. The median log(L$_{TIR}$/L$\arcmin_{HCN}$) values for sources with L$_{TIR}>10^{4.5}$L$_{\sun}$ and with L$_{TIR}<10^{4.5}$L$_{\sun}$ are $\sim$3.70$\pm$0.25 and $\sim$3.42$\pm$0.48, respectively. The uncertainties are the standard deviation of the means. The median log(L$_{TIR}$/L$\arcmin_{CS}$) values for sources with L$_{TIR}>10^{4.5}$L$_{\sun}$ and with L$_{TIR}<10^{4.5}$L$_{\sun}$ are $\sim$3.88$\pm$0.14 and $\sim$3.66$\pm$0.15, respectively. Therefore, considering uncertainties, there seems no threshold for HCN (4-3) and CS (7-6) lines in L$_{TIR}$--L$\arcmin_{mol}$ correlations in our sample. The correlations of L$_{TIR}$--L$\arcmin_{CS}$ and L$_{TIR}$--L$\arcmin_{HCN}$ are tight in the whole sample with L$_{TIR}$ down to at least $10^{3}$L$_{\sun}$. We also noticed that the sub-linear L$_{TIR}$--L$\arcmin_{HCN}$ correlation may even extend to the low luminosity (L$_{TIR}<10^{2}$L$_{\sun}$) end. However, we only have one data point with L$_{TIR}<10^{3}$L$_{\sun}$ in the L$_{TIR}$--L$\arcmin_{HCN}$ plot. More observations of low luminosity clumps would be helpful.

In panels (c) and (d) of Figure 5, we extend the L$_{TIR}$--L$\arcmin_{CS}$ and L$_{TIR}$--L$\arcmin_{HCN}$ correlations from Galactic molecular clumps to active galaxies. The data of active galaxies are from \cite{zhang14}. We only include their data with strong detections ($\geq4\sigma$).
Both L$_{TIR}$--L$\arcmin_{CS}$ and L$_{TIR}$--L$\arcmin_{HCN}$ correlations are very tight with slopes close to unit over about ten orders of magnitude in L$_{TIR}$. The best linear least-squares fits with uncertainties are listed
below, with R indicating the correlation coefficient:
\begin{equation*}
\begin{split}
\textrm{log}(L_{TIR})=(1.02\pm0.03)\times \textrm{log}(L\arcmin_{CS})\\+(3.80\pm0.04);~R=0.96\\
\textrm{log}(L_{TIR})=(1.03\pm0.02)\times \textrm{log}(L\arcmin_{HCN})\\+(3.58\pm0.04);~R=0.98
\end{split}
\end{equation*}

The slope ($1.02\pm0.03$) of CS (7-6) derived here is consistent with the one ($1.00\pm0.01$) in \cite{zhang14}. From Bayesian regression with LINMIX\_ERR (Kelly 2007), the slopes of L$_{TIR}$--L$\arcmin_{CS}$ and L$_{TIR}$--L$\arcmin_{HCN}$ are 1.05$\pm$0.02 and 1.03$\pm$0.02, respectively, which are consistent with results from linear least-squares fits. We noticed that the total IR luminosity in \cite{zhang14} was obtained by using the four IRAS bands equation as in other previous works \citep{gao04,wu05,wu10}, rather than by integrating the SED as we did toward massive clumps. However, in section 4.1, we will demonstrate that there is little difference between the total IR luminosity (L$_{IR}$) derived from IRAS four bands and the total IR luminosity (L$_{TIR}$) derived by integrating the SED. If we use L$_{IR}$ instead of L$_{TIR}$ for Galactic molecular clumps, the correlations for both Galactic molecular clumps and active galaxies in Figure 5 will become:
\begin{equation*}
\begin{split}
\textrm{log}(L_{IR})=(1.05\pm0.06)\times \textrm{log}(L\arcmin_{CS})\\+(3.85\pm0.04);~R=0.97\\
\textrm{log}(L_{IR})=(1.02\pm0.03)\times \textrm{log}(L\arcmin_{HCN})\\+(3.63\pm0.05);~R=0.99
\end{split}
\end{equation*} \\
These correlations are well consistent with those derived with L$_{TIR}$.

\section{Discussions}

\subsection{Caveats in estimating infrared luminosity and line luminosity}

\begin{figure*}
\centering
\includegraphics[angle=-90,scale=0.7]{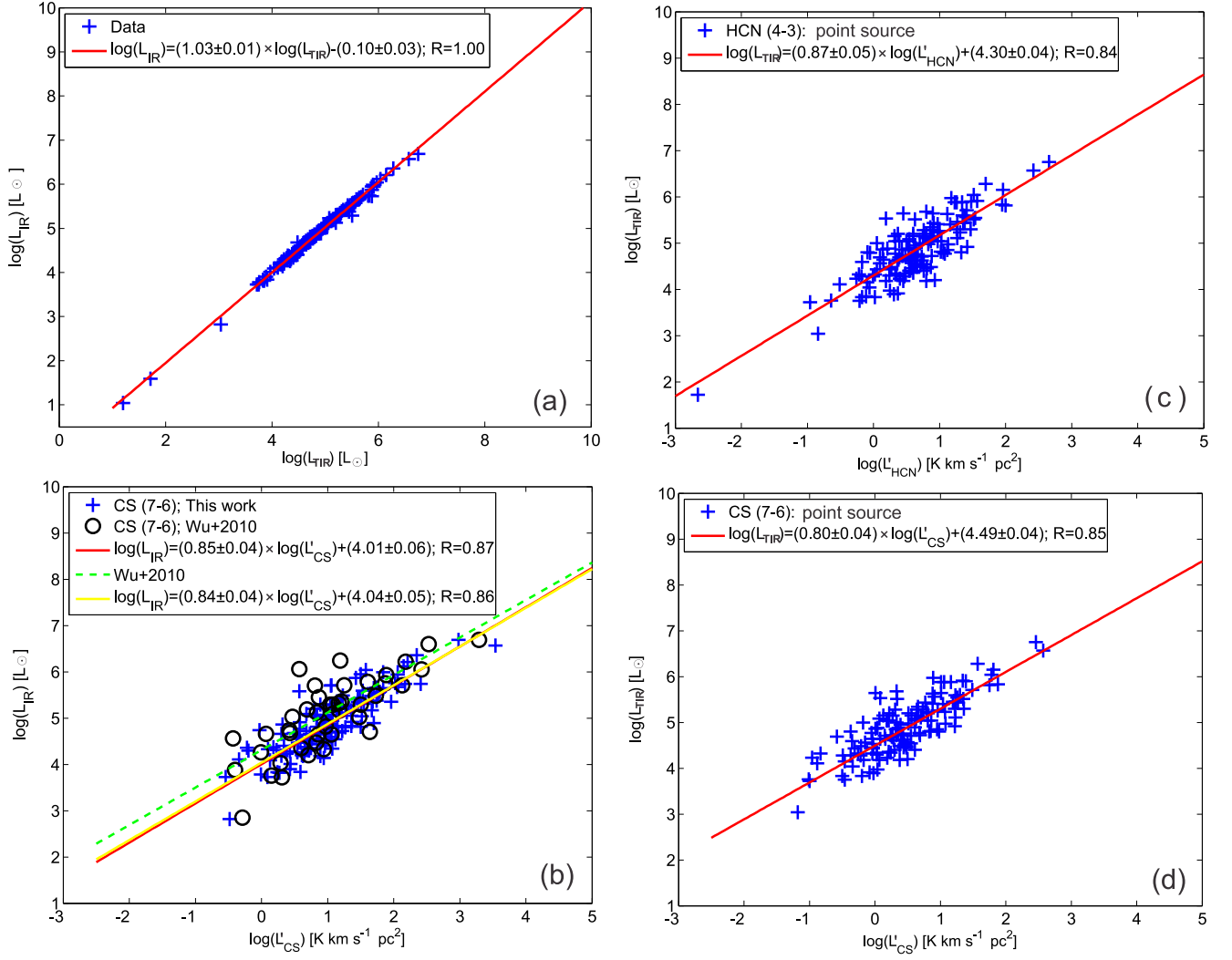}
\caption{Upper-left panel: correlation between IR luminosities (L$_{IR}$) derived from IRAS four bands and IR luminosities (L$_{TIR}$) derived from integrating whole SED. Upper-right panel: correlation between L$_{TIR}$ and beam-averaged L$\arcmin_{HCN}$. Lower-left panel: Comparison of L$_{IR}$--–L$\arcmin_{CS}$ correlations in our sample with correlations in the sample of Wu et al. (2010). Lower-right panel:  correlation between L$_{TIR}$ and beam-averaged L$\arcmin_{CS}$. }
\end{figure*}

\begin{figure*}[tbh!]
\centering
\includegraphics[angle=0,scale=0.85]{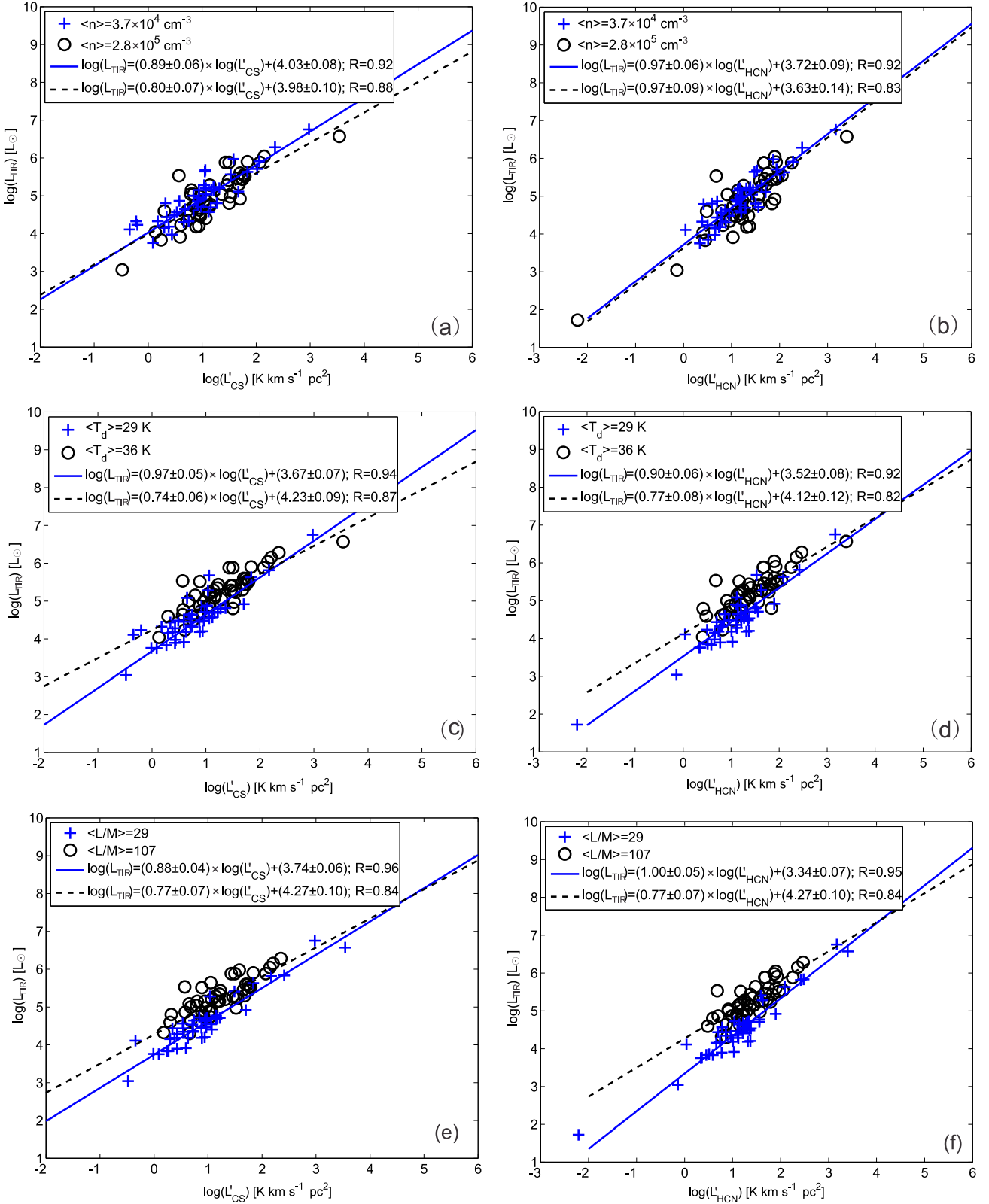}
\caption{Upper-left panel: L$_{TIR}$--–L$\arcmin_{CS}$ correlations for clumps with smallest volume densities (blue crosses) and for clumps with largest volume densities (open circles). Upper-right panel: L$_{TIR}$--–L$\arcmin_{HCN}$ correlations for clumps with smallest volume densities (blue crosses) and for clumps with largest volume densities (open circles). Middle-left panel: L$_{TIR}$--–L$\arcmin_{CS}$ correlations for clumps with smallest dust temperature (blue crosses) and for clumps with largest dust temperature (open circles). Middle-right panel: L$_{TIR}$--–L$\arcmin_{HCN}$ correlations for clumps with smallest dust temperature (blue crosses) and for clumps with largest dust temperature (open circles). Lower-left panel: L$_{TIR}$--–L$\arcmin_{CS}$ correlations for clumps with smallest luminosity-to-mass ratios (blue crosses) and for clumps with largest luminosity-to-mass ratios  (open circles). Lower-right panel: L$_{TIR}$--–L$\arcmin_{HCN}$ correlations for clumps with smallest luminosity-to-mass ratios  (blue crosses) and for clumps with largest luminosity-to-mass ratios  (open circles). }
\end{figure*}

\begin{figure*}
\centering
\includegraphics[angle=-90,scale=0.7]{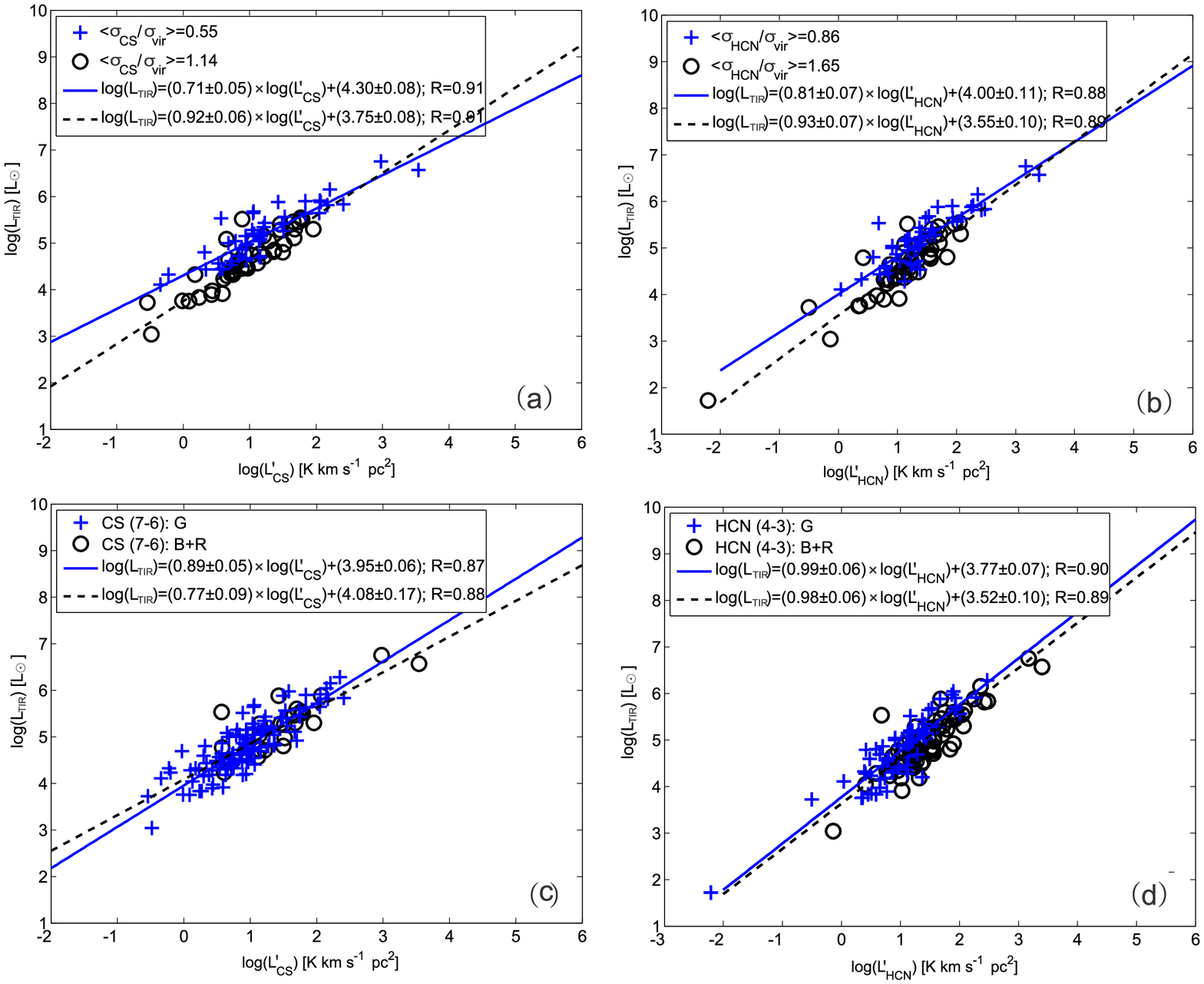}
\caption{Upper-left panel: L$_{TIR}$--–L$\arcmin_{CS}$ correlations for clumps with smallest $\sigma_{CS}/\sigma_{vir}$ (blue crosses) and for clumps with largest $\sigma_{CS}/\sigma_{vir}$ (open circles). Upper-right panel: L$_{TIR}$--–L$\arcmin_{HCN}$ correlations for clumps with smallest $\sigma_{HCN}/\sigma_{vir}$ (blue crosses) and for clumps with largest $\sigma_{HCN}/\sigma_{vir}$ (open circles). Lower-left panel: L$_{TIR}$--–L$\arcmin_{CS}$ correlations for clumps with symmetric profiles (blue crosses) and for clumps with asymmetric profiles (open circles). Upper-right panel:  L$_{TIR}$--–L$\arcmin_{HCN}$ correlations for clumps with symmetric profiles (blue crosses) and for clumps with asymmetric profiles (open circles).}
\end{figure*}

As shown in panel (a) of Figure 5, the correlation between infrared luminosity and CS (7-6) line luminosity derived for our sample is slightly different from \cite{wu10}. Such difference comes from the different methods used to derive infrared luminosity and CS (7-6) line luminosity. In our work, we used the total infrared luminosity (L$_{TIR}$) from integrating the whole SED \citep{fa04}. While in \cite{wu05,wu10}, the total infrared luminosity (L$_{IR}$ in L$_{\sun}$) was calculated based on the four IRAS bands:
\begin{equation}
L_{IR}=0.56\times D^{2}\times(13.48\times f_{12}+5.16\times f_{25}+2.58\times f_{60}+f_{100})
\end{equation}
where $f_{x}$ is the flux in band x in units of Jy, D is distance in kpc. We also derived L$_{IR}$ based on the four IRAS bands and present them in second column of Table 2. In panel (a) of Figure 7, we present the derived L$_{IR}$ and L$_{TIR}$. L$_{IR}$ and L$_{TIR}$ are nearly linearly correlated with each other as:
\begin{equation}
\textrm{log}(L_{IR})=(1.03\pm0.01)\times \textrm{log}(L_{TIR})-(0.10\pm0.03);~R=1.00
\end{equation}
In panel (b) of Figure 7, we investigate the relation between L$_{IR}$ and L$\arcmin_{CS}$. The L$_{IR}$--L$\arcmin_{CS}$ correlation for our sample is nearly the same as L$_{TIR}$--L$\arcmin_{CS}$ correlation, indicating that different methods in deriving infrared luminosity should not affect the interpretation of the Kennicutt-Schmidt law.

The intercept for our sample is smaller than the one of \cite{wu10}. It is probably because we used continuum size to calculate CS (7-6) line luminosity. Since lines are easily excited in subthermally populated
gas with densities more than an order of magnitude lower than critical density n$_{crit}$, the effective excitation density n$_{eff}$, rather than
n$_{crit}$, is more suitable to characterize the environment where a
transition is excited \citep{reiter11}. The CS (7-6) has a much higher effective excitation density ($2.1\times10^{6}$ cm$^{-3}$ at 20 K) than the median volume density ($\sim1.0\times10^{5}$ cm$^{-3}$) of our sample. Therefore the emission size of CS (7-6) should be smaller than the continuum emission size.

\cite{wu10} mapped 52 clumps in CS (7-6). The 25 of these 52 clumps were also mapped in 350 $\micron$ continuum emission by \cite{mue02}. The median ratio of CS (7-6) emission size (Size$_{cs}$) to the 350 $\micron$ continuum emission size (Size$_{0.35mm}$) is $\sim$0.87. Most sources both in our sample and the sample of \cite{wu10} are UC H{\sc ii} regions. The median infrared luminosity in the sample of \cite{wu10} is $1.06\times10^5$ L$_{\sun}$, which is comparable to the median infrared luminosity ($6.7\times10^4$ L$_{\sun}$) of our sample. Especially, the 114
most luminous clumps in our sample also has a median infrared luminosity of $1.1\times10^5$ L$_{\sun}$. Therefore, we assume that the sources in our sample share similar properties as the sources in \cite{wu10}. If we assume that the 1.2 mm continuum emission size (Size$_{1.2mm}$) is similar to Size$_{0.35mm}$ and all the sources have a constant Size$_{cs}$/Size$_{1.2mm}$ ratio of 0.87, we can estimate the overestimation factor in CS (7-6) line luminosity. We admit that further mapping observations are needed to test these assumptions. The 1.2 mm continuum angular sizes range from 26$\arcsec$ to 70$\arcsec$ in our sample. Therefore, the corresponding CS (7-6) line luminosity derived in section 3.2 should be overestimated by a factor of $\frac{Size_{1.2mm}^2+22^2}{(Size_{1.2mm}\times0.87)^2+22^2}-1$, which ranges from 17\% to 28\%. The overestimation of CS (7-6) line luminosity leads to a smaller intercept of L$_{IR}$--L$\arcmin_{CS}$ correlation. However, since the factor of overestimation for sources with different angular sizes only varies by $<$11\%, the slopes of correlations should not be severely affected by size assumption. Indeed, as shown in panel (a) of Figure 5, the slope of L$_{IR}$--L$\arcmin_{CS}$ correlation derived from our single pointing observations is very consistent with the slope derived from the mapping observations in \cite{wu10}.

\cite{wu10} only fitted their data for clumps with L$_{IR}>10^{4.5}$ L$_{\sun}$. The best linear least-squares fit of L$_{IR}$--L$\arcmin_{CS}$ for all the data of \cite{wu10} is:
\begin{equation*}
\textrm{log}(L_{IR})=(0.82\pm0.10)\times \textrm{log}(L\arcmin_{CS})+(4.23\pm0.13);~R=0.79\\
\end{equation*}
The best linear least-squares fit of L$_{IR}$--L$\arcmin_{CS}$ for our sample is:
\begin{equation*}
\textrm{log}(L_{IR})=(0.85\pm0.04)\times \textrm{log}(L\arcmin_{CS})+(4.01\pm0.06);~R=0.87\\
\end{equation*}
The best linear least-squares fit of L$_{IR}$--L$\arcmin_{CS}$ for both our sample and the sample of \cite{wu10} is:
\begin{equation*}
\textrm{log}(L_{IR})=(0.84\pm0.04)\times \textrm{log}(L\arcmin_{CS})+(4.04\pm0.05);~R=0.86\\
\end{equation*}

The slopes of these three correlations are consistent with each other considering errors, strongly indicating that the
slopes are not affected by the assuming of the same size
between dense gas and dust. We noticed that the effective excitation density ($2.4\times10^{5}$ cm$^{-3}$ at 20 K) of HCN (4-3) is comparable to the median volume density of our sample, indicating that the HCN (4-3) line luminosity should not be overestimated as severely as CS (7-6).

To further investigate the effect of size assumption on L$_{TIR}$-L$\arcmin_{mol}$ correlations, we derive a lower limit for line luminosity with Equation (1) assuming that the source size is much smaller than the beam size ($\theta_{s}\ll\theta_{beam}$). In panels (c) and (d) of Figure 7, we investigate the correlations between infrared luminosity and line luminosity derived under such point source assumption for HCN (4-3) and CS (7-6), respectively. We find that the correlations have similar slopes and slightly larger intercepts when compared with the values derived by assuming a source size the same as continuum emission size. The slope (0.80$\pm$0.04) for CS (7-6) is consistent with the value of \cite{wu10}. But the intercept (4.49$\pm$0.04) for CS (7-6) is slightly larger than the value of \cite{wu10}. Therefore, the line luminosities obtained in section 3.2 should be systematically overestimated especially for CS (7-6). However, the slopes of L$_{TIR}$--L$\arcmin_{mol}$ correlations are not affected as much as intercepts.

In conclusion, our single pointing results are consistent with previous mapping results considering the uncertainties. The slopes of L$_{TIR}$--L$\arcmin_{mol}$ correlations are not affected by uncertainties in measurements of L$_{TIR}$ and L$\arcmin_{mol}$.

\subsection{Dependance of dense molecular gas star formation law on physical parameters}

In Figure 8 and 9, we investigate how the dense molecular gas star formation law depends on different physical parameters, such as volume density, dust temperature, luminosity-to-mass ratio, $\sigma_{line}/\sigma_{vir}$, and line profiles for Galactic clumps. Taking ``volume density" for example, we firstly sorted the clumps according to their volume density values. Then we investigated the L$_{TIR}$--L$\arcmin_{CS}$ and L$_{TIR}$--L$\arcmin_{HCN}$ correlations toward the sources with volume densities ranked in the lower third and upper third, respectively. The median values of volume density in the lower third and the upper third groups are $3.7\times10^4$ cm$^{-3}$ and $2.8\times10^5$ cm$^{-3}$, respectively. As shown in panels (a) and (b) in Figure 8, the L$_{TIR}$--L$\arcmin_{CS}$ and L$_{TIR}$--L$\arcmin_{HCN}$ correlations seem not to depend on volume density. Especially for HCN (4-3), correlations of both low density and high density clumps are very close to linear. We notice that the volume density is anti-correlated with distance due to the different effective linear resolution. The determinations of some quantities, such as surface and volume densities for distant clumps, could be biased to lower values due to worse effective linear resolution \citep{wu10}. Therefore, that L$_{TIR}$--L$\arcmin_{mol}$ correlation does not change with volume density does not mean that star formation efficiencies (SFEs) are not affected by density. In other words, L$_{TIR}$--L$\arcmin_{mol}$ correlation is not biased by distance.

We also sorted the sample according to values of other physical parameters (dust temperature, luminosity-to-mass ratio, $\sigma_{line}/\sigma_{vir}$), and then investigated the L$_{TIR}$--L$\arcmin_{CS}$ and L$_{TIR}$--L$\arcmin_{HCN}$ correlations in the lower third and the upper third groups. The slopes ($\alpha$), intercepts ($\beta$) and correlation coefficients (R) of the correlations in each group are summarized in Table 3. In the first column of Table 3, we present the median values of the physical parameters of each group. Interestingly, we find that L$_{TIR}$--L$\arcmin_{CS}$ and L$_{TIR}$--L$\arcmin_{HCN}$ correlations show significant changes according to different dust temperature, luminosity-to-mass ratio and $\sigma_{line}/\sigma_{vir}$. The L$_{TIR}$--L$\arcmin_{mol}$ correlations show clear bimodal behavior. The correlations of clumps with lower dust temperature, lower luminosity-to-mass ratio and higher $\sigma_{line}/\sigma_{vir}$ are closer to linear. While the correlations of clumps with higher dust temperature, higher luminosity-to-mass ratio and lower $\sigma_{line}/\sigma_{vir}$ tend toward sublinear. It seems that sources with higher temperature and luminosity-to-mass ratios have higher $\frac{L_{TIR}}{L\arcmin_{mol}}$. Sources with smaller $\sigma_{line}/\sigma_{vir}$ ratios, which are more likely gravitationally bound, have higher $\frac{L_{TIR}}{L\arcmin_{mol}}$. Those sources with higher $\sigma_{line}/\sigma_{vir}$ ratios are more turbulent and probably are more affected by stellar feedback (like outflows), indicating that turbulence and stellar feedback could greatly reduce star formation efficiencies in molecular clouds, leading to a smaller $\frac{L_{TIR}}{L\arcmin_{mol}}$.

More evolved sources have higher infrared luminosities and probably have also consumed more gas, leading to higher luminosity-to-mass ratios \citep{ma13}. The luminosity-to-mass ratio should be a good evolutionary tracer \citep{ma13,liu13}. Therefore, the bimodal behavior of star formation law described above is more likely due to evolutionary effect. The sources in our sample were selected with IR colors similar to UC H{\sc ii} regions. They represent massive molecular clumps with embedded massive star formation. In contrast, massive infrared dark clouds (IRDCs) have very low luminosity-to-mass ratios since they have not yet formed massive stars. Furthermore, classical H{\sc ii} regions are luminous, but have very little associated molecular gas. Therefore, for sources at different evolutionary stages, the current (or observed) star formation efficiencies probed by L$_{IR}$/M$_{clump}$ or L$_{IR}$/L$\arcmin_{mol}$ should be very diverse at clump scale as also suggested by \cite{step16}. Our results in Figure 8 indicate that the slopes of L$_{TIR}$--L$\arcmin_{mol}$ correlations (or current star formation law) become more shallow as clumps evolve.

Is the bimodal behavior found in the L$_{TIR}$--L$\arcmin_{mol}$ correlations affected by the assuming of the same size between dense
gas (size$_{mol}$) and dust (size$_{dust}$). More evolved (higher L/M) sources likely also have higher temperature, which could be more efficient to excite the
dense gas tracer lines comparing to the cooler sources in the earlier stages \citep{moli16}. So
the late stage sources might have higher size$_{mol}$/size$_{dust}$ ratios than the
sources in the early stages. We investigate this effect with HCN (4-3). In extreme case, the line luminosity of a point source will be overestimated by a factor of $(35^2+22^2)/22^2=3.5$ if we take the median continuum size of 35$\arcsec$ as the line emission size.
If all the sources in the low L/M ($<L/M>=29$) subsample are point source like in HCN (4-3) emission, the HCN (4-3) line luminosity in panel (f) of Figure 8 should be systematically overestimated by a factor of $\sim$3.5 and the corresponding intercept of L$_{TIR}$--L$\arcmin_{mol}$ should be underestimated by $\sim$0.55. However, even in this extreme case, the intercept ($\sim3.89$) of the low L/M ($<L/M>=29$ subsample is still lower than that ($\sim4.27$) of the high L/M ($<L/M>=107$ subsample. Therefore, we argue that the bimodal behavior in the L$_{TIR}$--L$\arcmin_{mol}$ correlations is not greatly affected by size assumption. Future mapping observations will test this hypotheses.

In panels (c) and (d) of Figure 9, we investigate the L$_{TIR}$--L$\arcmin_{CS}$ and L$_{TIR}$--L$\arcmin_{HCN}$ correlations for sources with symmetric line profiles (``G") and asymmetric line profiles (``B+R"). In general, considering the errors, the correlations for sources with different line profiles do not show significant differences.

\subsection{Comparison between different tracers}

\begin{figure}
\centering
\includegraphics[angle=0,scale=0.47]{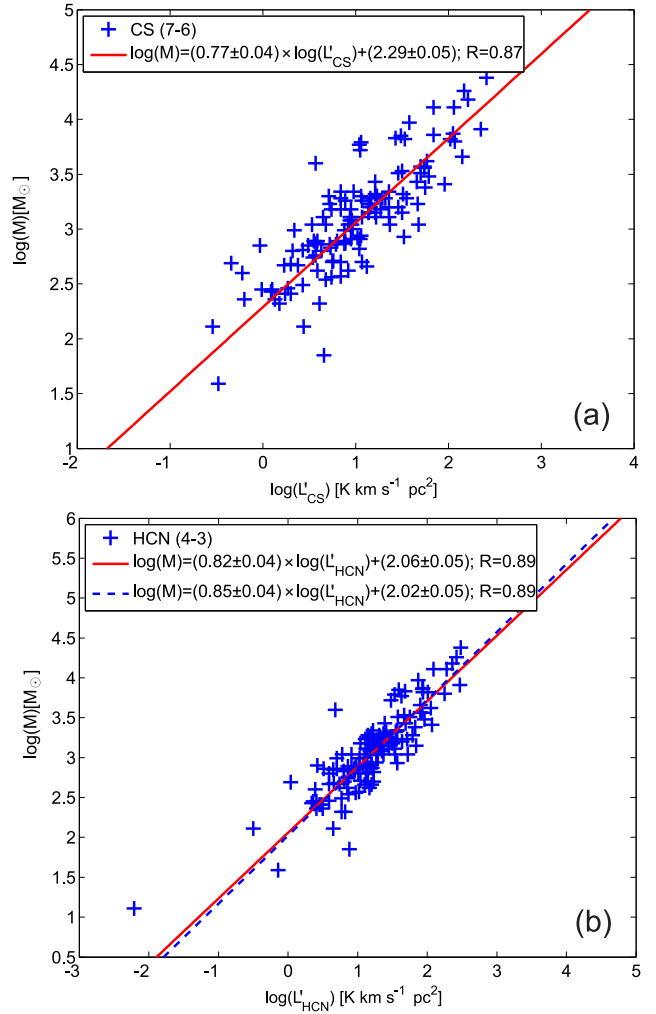}
\caption{Upper panel: Clump masses vs. luminosities of CS (7-6). Lower panels: Clump masses vs. luminosities of HCN (4-3). The red line represent least squares fits for the whole sample. The blue line represents least squares fits for clumps with infrared luminosities larger than 10$^{3}$ L$_{\sun}$. }
\end{figure}

\begin{figure*}
\centering
\includegraphics[angle=0,scale=0.85]{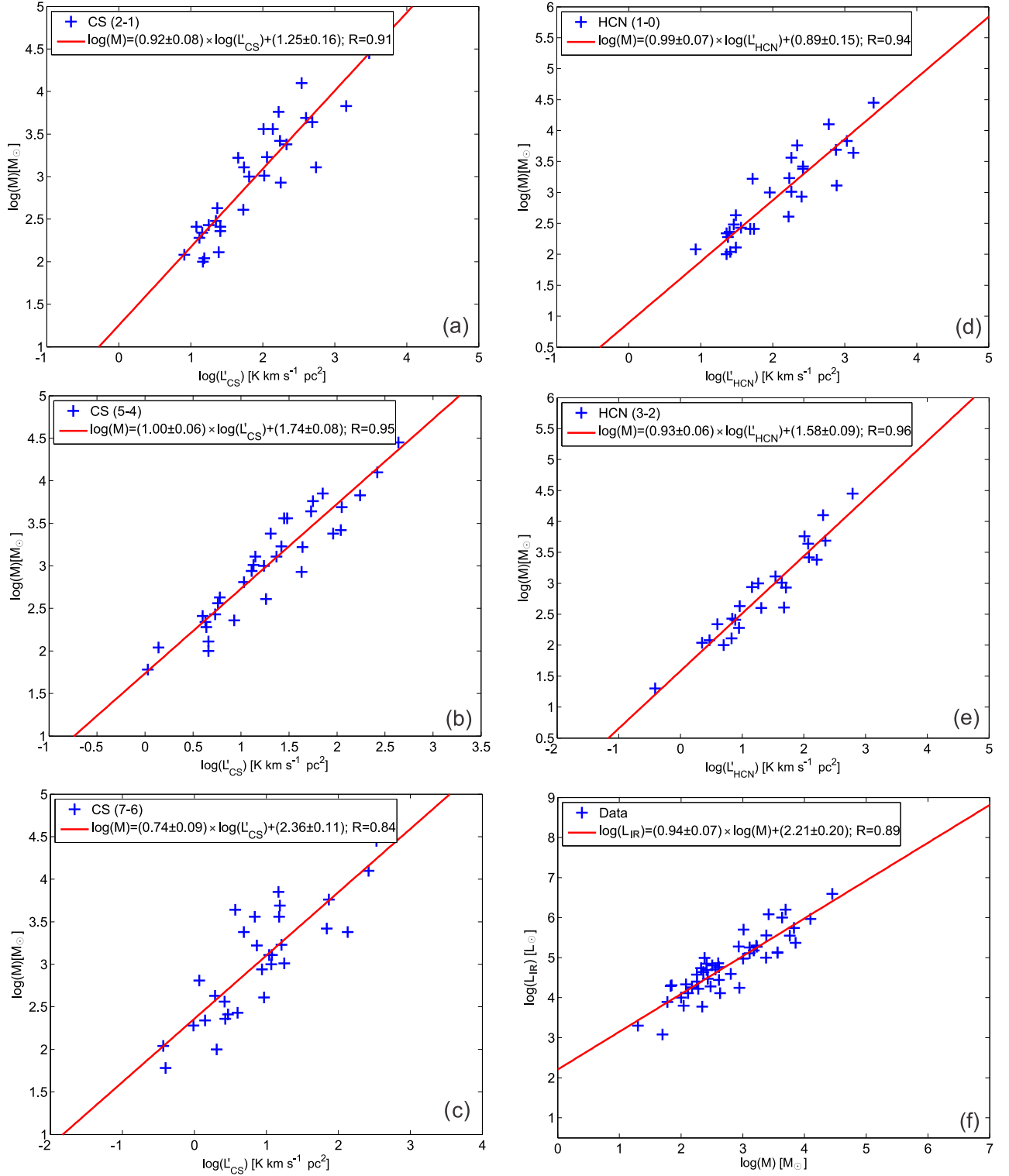}
\caption{Correlations for the sample of \cite{wu10}. (a). clump mass vs. L$\arcmin_{CS}$  for CS (2-1) data. (b). clump mass vs. L$\arcmin_{CS}$ for CS (5-4) data. (c). clump mass vs. L$\arcmin_{CS}$ for CS (7-6) data. (d). clump mass vs. L$\arcmin_{HCN}$ for HCN (1-0) data. (e). clump mass vs. L$\arcmin_{HCN}$ for HCN (3-2) data. (f). infrared luminosity vs. clump mass }
\end{figure*}

\begin{figure*}
\centering
\includegraphics[angle=-90,scale=0.6]{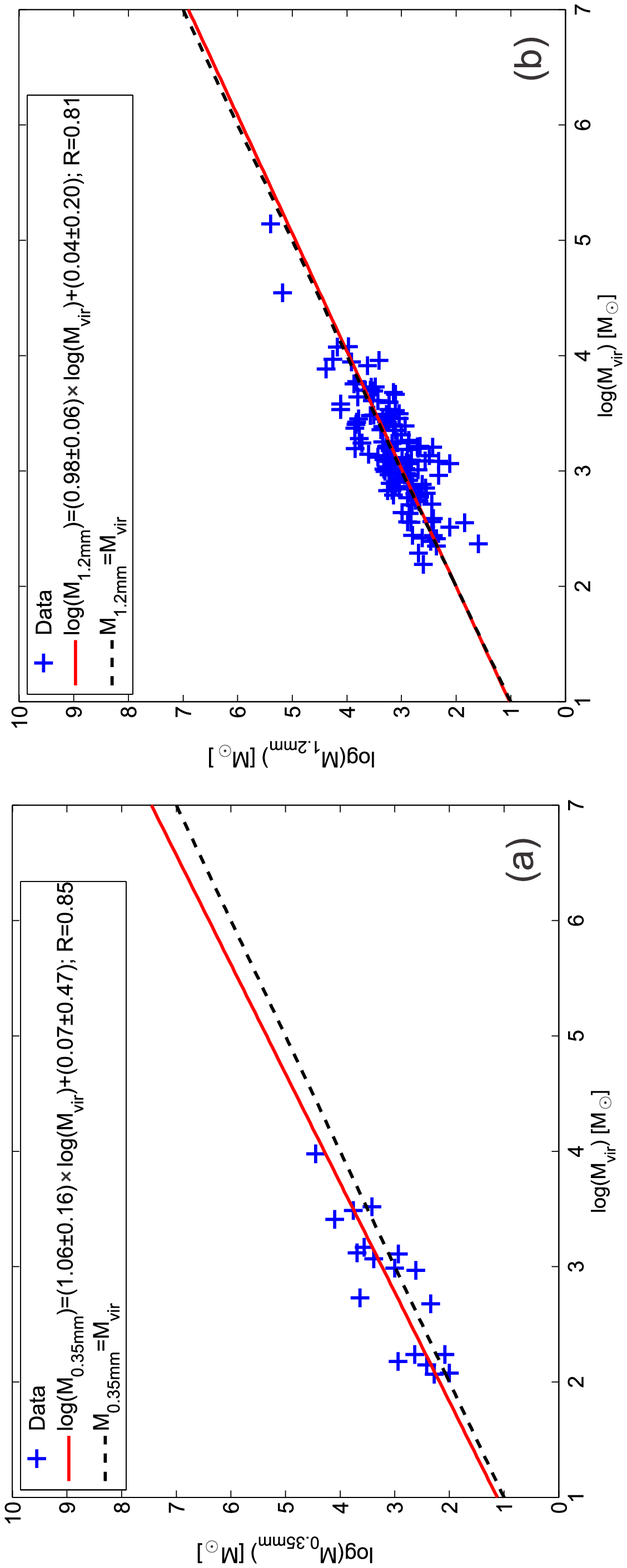}
\caption{(a). clump mass from \citep{mue02} vs. virial mass derived from HCN (3-2) correlations for the sample of \cite{wu10}. (b). clump mass derived from 1.1 mm continuum vs. virial mass derived from CS (7-6) correlations for our sample. The red lines represent the best power law fits. The black dashed lines represent linear correlations.}
\end{figure*}

In Figure 10, we present the correlations between clump masses and line luminosities for HCN (4-3) and CS (7-6). The M$_{clump}$--L$\arcmin_{CS}$ and M$_{clump}$--L$\arcmin_{HCN}$ correlations are very tight and sublinear. For comparison, we also investigate the correlations between clump masses and line luminosities for J=2-1, J=5-4, J=7-6 of CS and J=1-0, J=3-2 of HCN. The line luminosities are from \cite{wu10}. The clump masses were calculated with 350 $\micron$ continuum emission \citep{mue02}. As shown in Figure 11, the clump masses are also strongly correlated with line luminosities of these transitions. The slopes ($\alpha$), intercepts ($\beta$) and correlation coefficients (R) of the correlations were summarized in Table 4. For CS (7-6), the correlations for the sample of \cite{wu10} and our sample are consistent with similar slopes. In general, considering the errors, the correlations for lower transitions like J=2-1, J=5-4 of CS and J=1-0, J=3-2 of HCN are close to linear. While for the other higher transitions, the correlations are sublinear.

In Table 4, we also present the parameters of the L$_{TIR}$–-L$\arcmin_{mol}$ correlations. The parameters of the L$_{TIR}$--L$\arcmin_{mol}$ correlations for J=2-1, J=5-4 of CS and J=1-0, J=3-2 of HCN were taken from \cite{wu10}. It should be noted that \cite{wu10} only fitted the correlations for clumps with infrared luminosities larger than 10$^{4.5}$ L$_{\sun}$, where the initial mass function is fully sampled. Therefore, to compare with their results, we also fitted the L$_{TIR}$--L$\arcmin_{mol}$ correlations of CS (7-6) and HCN (4-3) for clumps with infrared luminosities larger than 10$^{4.5}$ L$_{\sun}$ in our sample. The derived slopes of CS (7-6) and HCN (4-3) are 0.72(0.05) and 0.74(0.06), respectively, which are smaller than the corresponding slopes for all clumps with infrared luminosities larger than 10$^{3}$ L$_{\sun}$. In general, as shown in Table 4, the L$_{TIR}$--L$\arcmin_{mol}$ correlations steepen for line transitions with lower upper energies and lower effective excitation densities. Especially for HCN (4-3) and CS (7-6), their L$_{TIR}$--L$\arcmin_{mol}$ correlations are apparently sublinear. \cite{nara08} predicted similar trend for external galaxies. However, the slopes of HCN transitions at clump scale (this work) or at galaxy scale \citep{zhang14} are all much higher than those predicted by \cite{nara08}. \cite{nara08} argued that lines with high critical densities (n$_{crit}$) greater than the mean density of most of the emitting clouds in a galaxy will have only a small amount of thermalized gas, leading to shallow slopes for their SFR--L$\arcmin_{mol}$ (or L$_{TIR}$--L$\arcmin_{mol}$) correlations. However, lines can easily excited in subthermally populated gas with densities more than an order of magnitude lower than n$_{crit}$ \citep{reiter11}. In addition, L$_{TIR}$--L$\arcmin_{mol}$ correlations seem not to depend on densities as shown in panels (a) and (b) of Figure 8. Therefore, \cite{nara08} may underestimated the line luminosities for high critical density tracers, leading to the shallow slopes in corresponding SFR--L$\arcmin_{mol}$ correlations.

\cite{lada10} found that the star formation rate (SFR) in molecular clouds is linearly proportional to the cloud mass above a gas volume density threshold of $\sim$10$^{4}$ cm$^{-3}$. We define the dense gas above this volume density threshold as star forming gas, which are directly related to star formation. Due to variations of the atmosphere mimic emission from extended astronomical objects, the maps obtained by ground-based bolometer arrays intrinsically filter the large-scale diffuse gas and are most sensitive to the highest column densities \citep{csen16}. Since diffuse gas is mainly distributed on the surface of dense clumps and is more likely to be filtered in bolometer observations, the continuum emission obtained by ground-based bolometer arrays is sensitive not only to high column density gas but also to high volume density gas if there are no overlapped clumps along the line of sight. In our sample, all the sources have single velocity component, indicating that there are no overlapped clumps along the same line of sight. The 1.2 mm continuum indeed traces high volume density (median value of $\sim1\times10^{5}$ cm$^{-3}$) star forming gas. The continuum emission obtained by
ground-based bolometer arrays is also often used to trace dense gas in other Galactic studies \citep{wien15,csen16}. In addition, the linear L$_{TIR}$–-M$_{clump}$ correlations in Figure 3 and panel (f) of Figure 11 also suggest that dust emission be one of the best mass tracers in star forming clumps.

For HCN (4-3) and CS (7-6),
the L$_{TIR}$–-L$\arcmin_{mol}$ correlations for clumps with infrared luminosities
larger than 10$^{4.5}$ L$_{\sun}$ are apparently sublinear.
What is the origin of such sublinear slopes? Since their upper energies and effective densities are much larger than the median dust temperature ($\sim$30 K) and median volume density (1$\times10^{5}$ cm$^{-3}$) of the whole sample, high effective excitation density tracers like HCN (4-3) and CS (7-6) may only trace the densest and warmest portions of star forming gas in clumps. It seems that high effective excitation density tracers cannot linearly trace the total masses of star forming gas, leading to a sublinear correlations for both M$_{clump}$--L$\arcmin_{mol}$ and L$_{TIR}$–-L$\arcmin_{mol}$ relations. The linear correlations of L$_{TIR}$–-L$\arcmin_{CS}$ and L$_{TIR}$–-L$\arcmin_{HCN}$ for both Galactic clumps and external galaxies presented in panels (c) and (d) of Figure 5 may be artificial caused by the large dynamical range of the data. Since L$_{TIR}$ is linearly correlated with M$_{clump}$, linear M$_{clump}$–-L$\arcmin_{mol}$ correlations for low effective excitation density tracers will naturally lead to linear L$_{TIR}$–-L$\arcmin_{mol}$ correlations. The different slopes of Kennicutt--Schmidt law for different molecular tracers are mainly determined by their excitation conditions. There should be no more underlying physics for different slopes indicated by various tracers.

We noticed that \cite{wu10} found nearly linear correlations between the dense gas luminosity (L$\arcmin_{mol}$) and
the virial mass (M$_{vir}$) even for high effective excitation density tracers like CS (7-6). Their results seem to differ from ours. However, \cite{wu10} derived the virial masses for the gas traced by CS (7-6) by using the linewidths of C$^{34}$S (5-4) lines and clump radii from CS (7-6) maps. However, since the high effective excitation density tracers like CS (7-6) and C$^{34}$S (5-4) only can trace the densest and warmest portions of dense gas in clumps, the virial masses derived from these lines cannot represent the total masses of all the star forming gas. If they use total clump mass instead of M$_{vir}$,
the corresponding L$\arcmin_{mol}$--M should be sublinear for high effective excitation density tracers like CS (7-6) as shown in panel (c) of Figure 11.  Low effective excitation density tracers can better trace the total star forming gas. As shown in panel (a) of Figure 12, the virial masses derived from HCN (3-2) is linearly correlated with clump masses derived from dust continuum emission, which can explain the linear correlation of M$_{clump}$--L$\arcmin_{HCN}$ in panel (e) of Figure 11. We derived the virial masses for the sources in our sample with the equation (4) in \cite{wu10}. We used the line widths of CS (7-6) and clump radii derived from 1.2 mm continuum. Interestingly, as shown in panel (b) of Figure 12, the clump masses derived from 1.2 mm continuum are linearly correlated with the virial masses, indicating that the 1.2 mm continuum emission well traces the bounded star forming gas. However, if we use the linewidths of optically thin lines (e.g. C$^{34}$S (5-4)) and the radii of actual CS (7-6) emission area as \cite{wu10} did, the corresponding virial masses should be smaller than the total clump masses. The use of virial mass is based on virial equilibrium. However, the status of massive clumps may deviate from virial equilibrium if they are affected by bulk motions like energetic outflows or infall. Due to the differences of excitation conditions, virial masses derived from different molecular transitions may vary very much \citep{wu10}. Therefore, we suggest dust continuum emission be a better tracer of the total star forming gas than virial masses.

\subsection{Connect star formation laws in Galactic clumps to external galaxies}

\cite{wu05} firstly tried to connect star formation laws in Galactic clumps to external galaxies. They found nearly linear L$_{IR}$--L$\arcmin_{mol}$ correlations for HCN (1-0) toward both Galactic dense clumps and galaxies. Previous works \citep{wu10,zhang14} and this work for other dense gas tracers also revealed linear L$_{IR}$--L$\arcmin_{mol}$ correlations when connecting Galactic dense clumps to external galaxies. \cite{wu05} argued that such linear correlations indicate a constant SFR per unit mass from the scale of dense clumps to that of distant galaxies \citep{wu05,wu10}. However, in extragalactic studies, one cannot easily resolve individual molecular clumps and distinguish between clouds at various evolutionary stages. As discussed in section 4.2, the current (or observed) star formation efficiencies probed by L$_{IR}$/M$_{clump}$ or L$_{IR}$/L$\arcmin_{mol}$ for clumps at different evolutionary stages are very diverse and the slopes of L$_{TIR}$--L$\arcmin_{mol}$ correlations become more shallow as clumps evolve. The global star formation law inferred in extragalactic studies is a mixture
of the star formation laws of different stages of star forming clumps/regions. Therefore, the linear L$_{IR}$--L$\arcmin_{mol}$ correlations may only indicate a constant star formation rate per unit mass (or L$_{IR}$/L$\arcmin_{mol}$) on scales much larger than clump scale. \cite{step16} suggested that L$_{IR}$/L$\arcmin_{mol}$ may become constant at some scales larger than $\sim$ 1 kpc because at only this size-scale will contain clumps to completely sample the clump mass function (CMF) and IMF.

In extragalactic studies, the Kennicutt--Schmidt scaling relations seem to change with redshifts, indicating that the star formation efficiency (or gas depletion time) also evolves with redshifts. For example, \cite{sant14}
found a molecular gas depletion time of 100--300 Myrs for a large sample of galaxies using dust continuum
measurements from Herschel and found an evolution of the molecular gas depletion time with redshift, by about a factor of 5 from $\textit{z}\sim0$ to $\textit{z}\sim5$. \cite{scov16} found a characteristic gas depletion time of 200--700 Myrs for a sample of galaxies at redshift $\textit{z}>1$. The entire population of star-forming galaxies at $\textit{z}>1$ has $\sim$2-5 times shorter gas depletion times than low-$\textit{z}$ galaxies \citep{scov16}. The gas depletion time of $\sim107^{+44}_{-31}$ Myr for Galactic massive clumps in our sample seems to be smaller than the characteristic gas depletion time in external galaxies. Interestingly, as discussed in section 3.4, the integrated Schmidt-Kennicutt laws for Galactic massive clumps can extend to high $\textit{z}$ star forming galaxies (SFGs). However dust continuum in extra-galactic studies traces both diffuse dust and dense dust. Therefore the extra-galactic studies can not seperate/resolve any dense gas phase from the total mass. The linear correlation of the integrated Schmidt-Kennicutt laws for both Galactic massive clumps and high $\textit{z}$ SFGs should indicate that the high $\textit{z}$ SFGs may have global properties (e.g., density, dense gas fraction, star formation efficiencies) at molecular cloud scale similar to Galactic massive clumps at clump scale.
In other words, the fraction of massive clumps (or dense gas) in molecular clouds of high $\textit{z}$ SFGs should be much higher than the Milky Way. While the low $\textit{z}$ galaxies have much lower star formation efficiencies than Galactic clumps, suggesting that their molecular clouds contain less massive clumps (or dense gas) than high $\textit{z}$ SFGs. In high $\textit{z}$ SFGs, dispersive gas motions (as opposed to ordered
rotation) and/or galaxy interactions will lead to compression in the highly dissipative ISM \citep{scov16}, enhancing the density and SFR per unit gas mass in their molecular clouds and creating more massive clumps (or dense gas) than low $\textit{z}$ galaxies.

\section{Summary}

We carried out HCN (4-3) and CS (7-6) observations, using the ASTE telescope, towards the 146 IRAS sources reported by \cite{fa04} in order to investigate the Kennicutt--Schmidt law for Galactic clumps. Our main conclusions are:

(1). The star formation rates indicated by total infrared
luminosities (L$_{TIR}$) are linearly correlated with
clump masses for those clumps with L$_{TIR}>10^{3}$ L$_{\sun}$, leading
to a constant gas depletion time of 107$^{+44}_{-31}$ Myr. The integrated Schmidt-Kennicutt laws for Galactic massive clumps can extend to high $\textit{z}$ star forming galaxies (SFGs). While low $\textit{z}$ galaxies have much smaller global star formation efficiencies (or longer gas depletion time) than Galactic massive clumps and high $\textit{z}$ SFGs. High $\textit{z}$ SFGs may have much larger fraction of massive clumps (or dense gas) in molecular clouds than low $\textit{z}$ galaxies.

(2). The L$_{TIR}$–-L$\arcmin_{mol}$ correlations for HCN (4-3) and
CS (7-6) for clumps are tight and sublinear over about
four orders of magnitude in L$_{TIR}$. We did not notice any
breakup of correlations at a luminosity threshold of 10$^{4.5}$ L$_{\sun}$ as suggested by Wu et al. (2010). Instead, the correlations
can extend down to clumps with  L$_{TIR}<10^{3}$ L$_{\sun}$. These correlations become linear when taking into
account external galaxies.

(3). We find that L$_{TIR}$–-L$\arcmin_{CS}$ and L$_{TIR}$–-L$\arcmin_{HCN}$ correlations
for clumps with lower dust temperature, lower
luminosity-to-mass ratio and higher $\sigma_{line}/\sigma_{vir}$ are closer
to linear. While the correlations for clumps with higher
dust temperature, higher luminosity-to-mass ratio and
lower $\sigma_{line}/\sigma_{vir}$ tend toward sublinear. Such bimodal
behavior of L$_{TIR}$–-L$\arcmin_{mol}$ correlations may be due to evolutionary
effects. The slopes of L$_{IR}$--L$\arcmin_{mol}$ correlations (or current star formation law) become more shallow as clumps evolve.

(4). The line luminosities of molecular tracers with upper
energies and effective excitation densities lower than the median
dust temperature and median volume density of the
whole sample are linearly correlated with clump masses.
While high effective excitation density tracers cannot linearly trace the total clump masses, leading to a sublinear correlations for both M$_{clump}$--L$\arcmin_{mol}$ and L$_{TIR}$–-L$\arcmin_{mol}$ relations.
Since clump masses traced by dust continuum
emission are linearly correlated with infrared luminosities.
The (sub)linear M$_{clump}$–-L$\arcmin_{mol}$ correlations can naturally
explain the (sub)linear L$_{TIR}$–-L$\arcmin_{mol}$ correlations (i.e. the
Kennicutt--Schmidt law) for different molecular line tracers.

\section*{Acknowledgment}
\begin{acknowledgements}

We are grateful to the ASTE staff. Tie Liu is supported by KASI fellowship. Y. Wu is partly supported by the China Ministry of Science and
Technology under State Key Development Program for Basic Research (No.2012CB821800), the grants of NSFC No.11373009 and No.11433008. Ke Wang acknowledge the support from ESO fellowship and DFG Priority Program 1573 (``Physics of the Interstellar Medium'') grant WA3628-1/1. This work was carried out in part at the Jet Propulsion Laboratory, operated for NASA by the California Institute of Technology. S. L. Qin is supported by NSFC  under grant No. 11373026, and Top Talents Program of Yunnan Province (2015HA030). MJ acknowledges the support of the Academy of Finland Grant No. 1285769. LB acknowledges support from CONICYT grant PFB-06. J.-E.L. was supported by the Basic Science Research Program through the National Research Foundation of Korea (NRF) (grant No. NRF-2015R1A2A2A01004769) and the Korea Astronomy and Space Science Institute under the R\&D program (Project No. 2015–1-320–18) supervised by the Ministry of Science, ICT, and Future Planning. The anonymous referee provided very insightful comments.

\end{acknowledgements}

\clearpage

\begin{longtable*}{cccccccccccccc}
\tablecolumns{12} \tablewidth{0pc}
\tablecaption{Parameters of molecular lines in Single-dish observations} \tablehead{
\colhead{IRAS} &  & HCN (4-3) & & & & && CS (7-6)  &\\
\cline{2-6}\cline{8-12}\\
 & \colhead{$\int T_{A}^{*}dV$} & \colhead{V$_{lsr}$} & \colhead{FWHM } & \colhead{T$_{A}^{*}$ }  & \colhead{Profile } &
  & \colhead{$\int T_{A}^{*}dV$} & \colhead{V$_{lsr}$} & \colhead{FWHM } & \colhead{T$_{A}^{*}$ } & \colhead{Profile }
 \\
\colhead{}  & \colhead{(K~km~s$^{-1}$)}    &\colhead{(km~s$^{-1}$)} &
\colhead{(km~s$^{-1}$)} &
\colhead{(K)} & & & \colhead{(K~km~s$^{-1}$)}    &\colhead{(km~s$^{-1}$)} &
\colhead{(km~s$^{-1}$)} &
\colhead{(K)}
  } 	
08076-3556	&   --         &    --        &     --      &   --   & --& &    --           &   --         &  --        & --   & --   \\
08303-4303	&	  9.14(0.29)& 	15.22(0.09)  &	5.70(0.22)  & 	1.51	& G& & 	 5.61(0.17)  &	 15.00(0.05) &	3.38(0.12) &	1.56	& G   \\
08448-4343	&	 13.84(0.28)& 	 3.41(0.06)  &	6.24(0.15)  & 	2.08	& R& & 	 6.24(0.17)  &	  3.19(0.05) &	3.86(0.13) &	1.52	& G   \\
08470-4243	&	  9.16(0.36)& 	13.46(0.10)  &	5.10(0.26)  & 	1.69	& B& & 	 4.79(0.17)  &	 12.73(0.04) &	2.60(0.12) &	1.73	& G   \\
09002-4732	&	 21.87(0.25)& 	 3.93(0.03)  &	5.02(0.07)  & 	4.10	& G& & 	14.54(0.15)  &	  3.20(0.02) &	3.32(0.04) &	4.11	& G   \\
09018-4816	&	 20.76(1.08)& 	10.77(0.07)  &	6.27(0.27)  & 	3.11	& B& & 	14.88(0.18)  &	 10.28(0.03) &	5.32(0.07) &	2.63	& G   \\
09094-4803	&     --       & 	    --      &     --      &   --   & --&   &      --        &       --     &      --    &       & --   \\
10365-5803	&		7.84(0.30)&  -18.94(0.08) &	5.20(0.25)  & 	1.42	& B&   & 	 3.44(0.39)  &	-19.32(0.19) &	3.58(0.34) &	0.90	& G   \\
11298-6155	&	  4.38(0.17)& 	 33.04(0.09) &	4.74(0.22)  & 	0.87	& G& & 	 2.85(0.16)  &	 32.74(0.08) &	3.05(0.21) &	0.88	& G   \\
11332-6258	&		1.41(0.21)&  -16.09(0.22) &	3.04(0.60)  & 	0.43	& G&   & 	 1.30(0.14)  &	-15.99(0.19) &	3.36(0.43) &	0.36	& G   \\
11590-6452	&   0.63(0.11)&   -3.94(0.16) &	1.50(0.37)  & 	0.40	& G&   & 	--           &    --        &     --     &  --   & --   \\
12320-6122	&	 17.62(0.47)&  -42.54(0.08) &	6.36(0.20)  & 	2.60	& G&   & 	10.95(0.22)  &	-43.10(0.03) &	3.65(0.09) &	2.82	& G   \\
12326-6245	&	 54.99(0.56)&  -39.13(0.04) &	9.26(0.12)  & 	5.58	& B&   & 	45.21(0.24)  &	-39.79(0.01) &	5.74(0.04) &	7.40	& G   \\
12383-6128	&		5.62(0.16)&  -37.94(0.05) &	3.55(0.12)  & 	1.49	& G&   & 	 4.19(0.14)  &	-38.75(0.04) &	2.43(0.09) &	1.62	& G   \\
12572-6316	&	  3.42(0.14)& 	 29.51(0.14) &	6.76(0.32)  & 	0.48	& G& & 	 1.91(0.11)  &	 29.77(0.15) &	5.59(0.40) &	0.32	& G   \\
13079-6218	&	 47.05(0.59)&  -40.51(0.06) & 11.27(0.25)  & 	3.92	& B&   & 	29.85(0.32)  &	-41.30(0.04) &	7.82(0.12) &	3.59	& G   \\
13080-6229	&	 17.06(0.31)&  -33.95(0.06) &	6.92(0.16)  & 	2.32	& G&   & 	 8.82(0.17)  &	-34.67(0.05) &	5.40(0.13) &	1.54	& G   \\
13111-6228	&	 17.91(0.24)&  -38.69(0.04) &	6.11(0.10)  & 	2.75	& G&   & 	14.46(0.14)  &	-39.38(0.02) &	4.20(0.05) &	3.23	& G   \\
13134-6242	&	 23.76(0.77)&  -31.58(0.11) &	8.72(0.20)  & 	2.56	& B&   & 	12.28(0.20)  &	-32.51(0.05) &	6.39(0.13) &	1.80	& G   \\
13140-6226	&		9.99(0.27)&  -32.73(0.09) &	7.31(0.26)  & 	1.28	& G&   & 	 4.59(0.22)  &	-33.92(0.10) &	5.36(0.38) &	0.80	& G   \\
13291-6229	&		6.45(0.31)&  -37.06(0.09) &	4.60(0.18)  & 	1.32	& B&   & 	 4.18(0.09)  &	-37.43(0.03) &	2.96(0.08) &	1.33	& G   \\
13291-6249	&	 20.41(0.13)&  -32.43(0.02) &	6.15(0.04)  & 	3.12	& G&   & 	11.85(0.13)  &	-32.93(0.02) &	4.16(0.06) &	2.68	& G   \\
13295-6152	&   1.79(0.16)&  -43.53(0.18) &	4.22(0.42)  & 	0.40	& G&   & 	 --          &     --       &    --      &        & --   \\
13471-6120	&	 13.12(0.17)&  -57.34(0.04) &	5.91(0.09)  & 	2.09	& G&   & 	 9.38(0.12)  &	-58.23(0.02) &	3.47(0.06) &	2.54	& G   \\
13484-6100	&	 20.38(0.61)&  -54.09(0.12) &	8.87(0.49)  & 	2.16	& R&   & 	13.82(0.29)  &	-54.83(0.06) &	6.04(0.17) &	2.15	& G   \\
14013-6105	&	 30.56(0.70)&  -54.57(0.02) &	5.24(0.08)  & 	5.48	& R&   & 	17.50(0.13)  &	-55.00(0.02) &	4.25(0.04) &	3.87	& G   \\
14050-6056	&		8.33(0.16)&  -48.10(0.05) &	5.70(0.12)  & 	1.37	& G&   & 	 6.73(0.10)  &	-48.61(0.02) &	2.91(0.05) &	2.17	& G   \\
14164-6028	&		6.83(0.21)&  -45.25(0.13) &	8.48(0.32)  & 	0.76	& G&   & 	 4.16(0.09)  &	-46.16(0.05) &	4.92(0.13) &	0.79	& G   \\
14206-6151	&		0.94(0.11)&  -48.77(0.15) &	2.65(0.35)  & 	0.33	& G&   & 	 0.40(0.11)  &	-49.60(0.23) &	1.67(0.65) &	0.22	& G   \\
14212-6131	&	 10.07(0.30)&  -49.25(0.09) &	6.95(0.28)  & 	1.36	& G&   & 	 3.40(0.22)  &	-49.65(0.19) &	6.66(0.57) &	0.48	& G   \\
14382-6017	&		7.76(0.40)&  -60.28(0.07) &	5.08(0.27)  & 	1.43	& B&   & 	 5.24(0.13)  &	-60.55(0.04) &	3.30(0.10) &	1.49	& G   \\
14453-5912	&	 10.26(0.47)&  -40.17(0.07) &	4.79(0.27)  & 	2.01	& B&   & 	 5.56(0.11)  &	-40.32(0.02) &	2.18(0.05) &	2.40	& G   \\
14498-5856	&	 21.88(0.36)&  -49.70(0.05) &	8.27(0.15)  & 	2.49	& B&   & 	13.34(0.20)  &	-50.03(0.03) &	3.96(0.08) &	3.17	& G   \\
15122-5801	&		3.78(0.14)&  -60.28(0.10) &	5.58(0.24)  & 	0.64	& G&   & 	 2.85(0.13)  &	-60.87(0.07) &	3.33(0.18) &	0.80	& G   \\
15254-5621	&	 17.50(0.20)&  -66.88(0.03) &	5.43(0.08)  & 	3.03	& G&   & 	10.28(0.15)  &	-67.94(0.03) &	3.78(0.07) &	2.56	& G   \\
15290-5546	&	 31.27(0.70)&  -88.48(0.11) & 10.68(0.21)  & 	2.75	& R&   & 	22.10(0.23)  &	-88.29(0.03) &	5.49(0.07) &	3.78	& G   \\
15384-5348	&		6.71(0.16)&  -40.60(0.05) &	4.53(0.12)  & 	1.39	& G&   & 	 3.94(0.11)  &	-41.37(0.04) &	3.02(0.10) &	1.23	& G   \\
15394-5358	&	 37.78(0.83)&  -40.70(0.06) &	8.23(0.17)  & 	4.31	& R&   & 	14.13(0.18)  &	-40.78(0.01) &	5.93(0.10) &	2.24	& G   \\
15408-5356	&	 52.13(0.96)&  -40.13(0.06) &	7.69(7.69)  & 	6.37	& R&   & 	34.08(1.16)  &	-41.07(0.08) &	6.83(0.12) &	4.69	& R   \\
15411-5352	&	 23.47(0.42)&  -40.98(0.05) &	5.32(0.09)  & 	4.15	& G&   & 	11.17(0.20)  &	-41.35(0.04) &	4.15(0.09) &	2.53	& G   \\
15437-5343	&   1.52(0.15)&  -83.81(0.24) &	4.69(0.54)  & 	0.31	& G&   & 	   --         &     --       &   --       &  --   & --   \\
15439-5449	&		9.12(0.34)&  -53.43(0.09) &	5.04(0.17)  & 	1.70	& R&   & 	 4.58(0.09)  &	-53.93(0.03) &	3.35(0.08) &	1.28	& G   \\
15502-5302	&	 25.44(0.26)&  -91.58(0.05) &	9.45(0.12)  & 	2.53	& G&   & 	17.19(0.20)  &	-92.55(0.04) &	7.04(0.11) &	2.29	& G   \\
15520-5234	&	101.90(0.61)&  -41.18(0.02) & 10.21(0.05)  & 	9.38	& B&   & 	92.63(0.26)  &	-41.25(0.01) &	8.42(0.03) &	10.33 & B   \\
15522-5411	&	  5.08(0.47)&  -46.83(0.15) &	4.03(0.42)  & 	1.18	& R&   & 	 4.45(0.21)  &	-47.07(0.08) &	3.60(0.21) &	1.16	& G   \\
15557-5215	&	 10.40(0.35)&  -67.58(0.16) & 10.58(0.48)  & 	0.92	& G&   & 	 4.30(0.24)  &	-68.08(0.19) &	7.51(0.58) &	0.54	& G   \\
15567-5236	&	 28.46(0.31)& -107.32(0.04) &	6.96(0.09)  & 	3.84	& G&   & 	23.25(0.22)  & -107.98(0.02) &	4.69(0.05) &	4.65	& G   \\
15570-5227	&		5.01(0.14)& -100.15(0.05) &	3.73(0.12)  & 	1.26	& G&   & 	 4.03(0.09)  & -100.58(0.03) &	2.46(0.07) &	1.53	& G   \\
15584-5247	&		2.63(0.18)&  -76.16(0.16) &	4.80(0.36)  & 	0.52	& G&   & 	 1.48(0.16)  &	-76.90(0.17) &	2.82(0.40) &	0.49	& G   \\
15596-5301	&	 22.83(1.73)&  -73.91(0.12) &	6.54(0.37)  & 	3.28	& B&   & 	15.56(0.26)  &	-74.44(0.04) &	5.41(0.11) &	2.70	& G   \\
16026-5035	&		4.24(0.25)&  -78.49(0.13) &	4.52(0.32)  & 	0.88	& G&   & 	 2.75(0.11)  &	-79.00(0.05) &	2.55(0.12) &	1.01	& G   \\
16037-5223	&	 14.48(0.32)&  -80.98(0.09) &	8.64(0.22)  & 	1.57	& B&   & 	13.93(0.24)  &	-80.88(0.06) &	6.54(0.13) &	2.00	& G   \\
16060-5146	&	 30.37(1.39)&  -88.79(0.10) &	9.60(0.43)  & 	2.97	& B&   & 	17.14(0.53)  &	-89.95(0.07) &	5.80(0.19) &	2.77	& B   \\
16065-5158	&	 54.32(1.18)&  -61.55(0.12) & 12.54(0.58)  & 	4.07	& R&   & 	50.47(0.35)  &	-62.25(0.03) &	8.45(0.07) &	5.61	& B   \\
16071-5142	&	 45.79(1.62)&  -85.91(0.18) & 11.01(0.54)  & 	3.91	& B&   & 	28.60(0.32)  &	-86.67(0.04) &	7.84(0.11) &	3.42	& G   \\
16076-5134	&	 50.50(0.49)&  -86.01(0.07) & 16.77(0.27)  & 	2.83	& R&   & 	39.74(0.54)  &	-87.32(0.06) &	9.51(0.25) &	3.92	& B   \\
16119-5048	&	 21.71(0.68)&  -48.42(0.11) &	9.15(0.38)  & 	2.23	& R&   & 	12.16(0.25)  &	-48.43(0.05) &	4.71(0.13) &	2.43	& G   \\
16132-5039	&		3.69(0.14)&  -47.01(0.06) &	3.51(0.16)  & 	0.99	& G&   & 	 1.95(0.09)  &	-47.59(0.05) &	1.99(0.11) &	0.92	& G   \\
16158-5055	&		9.40(0.23)&  -49.39(0.10) &	8.69(0.26)  & 	1.02	& G&   & 	 6.74(0.14)  &	-50.33(0.04) &	3.85(0.09) &	1.64	& G   \\
16164-5046	&	 76.99(1.60)&  -56.22(0.07) &	9.53(0.21)  & 	7.59	& R&   & 	67.31(0.41)  &	-57.40(0.02) &	7.76(0.06) &	8.15	& R   \\
16172-5028	&	 29.87(0.39)&  -52.06(0.06) &	8.92(0.13)  & 	3.15	& G&   & 	38.58(0.40)  &	-52.15(0.03) &	6.76(0.08) &	5.37	& G   \\
16177-5018	&		6.19(0.57)&  -50.93(0.14) &	6.47(0.48)  & 	0.90	& R&   & 	 4.80(0.41)  &	-51.60(0.11) &	5.03(0.37) &	0.90	& R   \\
16272-4837	&	 30.11(0.83)&  -45.79(0.11) &	9.06(0.72)  & 	3.12	& B&   & 	12.29(0.20)  &	-46.42(0.04) &	5.81(0.12) &	1.99	& G   \\
16297-4757	&	 10.53(0.24)&  -79.92(0.08) &	6.89(0.19)  & 	1.44	& G&   & 	12.69(0.16)  &	-79.72(0.03) &	5.68(0.09) &	2.10	& G   \\
16304-4710	&		1.81(0.12)&  -61.00(0.13) &	3.87(0.28)  & 	0.44	& G&   & 	 0.76(0.11)  &	-62.28(0.24) &	3.56(0.63) &	0.20	& G   \\
16313-4729	&		5.86(0.16)&  -72.29(0.08) &	6.09(0.20)  & 	0.90	& G&   & 	 3.00(0.16)  &	-73.20(0.18) &	7.06(0.46) &	0.40	& G   \\
16318-4724	&	 15.54(0.39)& -120.54(0.09) &	9.10(0.30)  & 	1.60	& R&   & 	11.67(0.17)  & -121.43(0.05) &	7.27(0.13) &	1.51	& G   \\
16330-4725	&	 19.56(0.17)&  -74.09(0.03) &	6.99(0.07)  & 	2.63	& G&   & 	14.60(0.13)  &	-74.62(0.02) &	5.70(0.06) &	2.41	& G   \\
16344-4658	&	 16.36(0.39)&  -48.85(0.10) & 12.41(0.30)  & 	1.24	& B&   & 	12.84(0.54)  &	-48.96(0.07) &	7.75(0.23) &	1.56	& B   \\
16348-4654	&	 31.52(0.50)&  -47.23(0.08) & 12.21(0.23)  & 	2.43	& B&   & 	17.65(0.21)  &	-47.92(0.04) &	7.32(0.11) &	2.27	& G   \\
16351-4722	&	 61.00(1.15)&  -40.30(0.06) &	9.10(0.20)  & 	6.29	& B&   & 	55.22(0.67)  &	-40.64(0.04) &	7.22(0.10) &	7.18	& B   \\
16362-4639	&		2.68(0.13)&  -37.38(0.12) &	5.07(0.31)  & 	0.50	& G&   & 	 0.53(0.07)  &	-37.93(0.12) &	1.85(0.30) &	0.27	& G   \\
16372-4545	&	 10.64(0.33)&  -57.54(0.06) &	5.69(0.18)  & 	1.76	& R&   & 	 8.72(0.26)  &	-58.44(0.06) &	4.48(0.11) &	1.83	& R   \\
16385-4619	&	 21.37(0.30)& -117.85(0.07) & 10.59(0.26)  & 	1.90	& B&   & 	19.40(0.16)  & -118.02(0.03) &	6.45(0.07) &	2.82	& G   \\
16424-4531	&	 17.39(0.30)&  -31.06(0.05) &	7.89(0.19)  & 	2.07	& G&   & 	 3.58(0.18)  &	-34.57(0.07) &	2.96(0.20) &	1.13	& G   \\
16445-4459	&	 10.37(0.23)& -121.43(0.07) &	6.92(0.22)  & 	1.41	& G&   & 	 4.55(0.15)  & -122.31(0.08) &	5.42(0.24) &	0.79	& G   \\
16458-4512	&	 17.40(0.36)&  -49.92(0.07) &	7.22(0.24)  & 	2.27	& R&   & 	 4.57(0.24)  &	-50.75(0.09) &	4.18(0.29) &	1.03	& G   \\
16484-4603	&	 25.05(0.62)&  -31.58(0.08) &	7.53(0.32)  & 	3.13	& R&   & 	16.30(0.23)  &	-32.19(0.03) &	4.76(0.11) &	3.21	& G   \\
16487-4423	&		6.76(0.17)&  -43.24(0.06) &	5.75(0.19)  & 	1.11	& R&   & 	 2.35(0.11)  &	-43.92(0.08) &	3.45(0.19) &	0.64	& G   \\
16489-4431	&		6.20(0.18)&  -40.40(0.07) &	5.06(0.19)  & 	1.15	& G&   & 	 2.75(0.11)  &	-40.74(0.06) &	3.18(0.16) &	0.81	& G   \\
16506-4512	&	 19.81(0.39)&  -25.80(0.03) &	4.31(0.07)  & 	4.31	& R&   & 	12.18(0.21)  &	-25.84(0.02) &	2.84(0.04) &	4.02	& R   \\
16524-4300	&	 19.11(0.38)&  -40.56(0.05) &	5.38(0.07)  & 	3.34	& R&   & 	 9.87(0.11)  &	-41.30(0.02) &	3.49(0.05) &	2.66	& G   \\
16547-4247	& 117.07(1.30)&  -29.73(0.08) & 17.16(0.31)  & 	6.41	& B&   & 	53.42(0.37)  &	-30.27(0.03) &	9.95(0.11) &	5.05	& B   \\
16562-3959	&	 71.50(0.48)&  -11.90(0.02) &	6.85(0.06)  & 	9.80	& R&   & 	38.38(0.22)  &	-12.53(0.01) &	4.55(0.03) &	7.93	& G   \\
16571-4029	&	 20.79(1.13)&  -14.69(0.12) &	5.65(0.41)  & 	3.46	& R&   & 	14.24(0.27)  &	-15.03(0.03) &	3.12(0.07) &	4.29	& G   \\
17006-4215	&	 15.92(0.26)&  -24.22(0.05) &	5.71(0.11)  & 	2.62	& G&   & 	13.40(0.13)  &	-24.69(0.01) &	3.06(0.04) &	4.12	& G   \\
17008-4040	&	 32.02(0.67)&  -15.93(0.07) &	6.67(0.12)  & 	4.51	& R&   & 	20.00(0.24)  &	-16.67(0.03) &	5.01(0.07) &	3.75	& G   \\
17016-4124	&	 83.23(1.44)&  -26.48(0.11) & 12.89(0.45)  & 	6.07	& R&   & 	32.83(0.42)  &	-26.80(0.05) &	9.12(0.15) &	3.38	& G   \\
17136-3617	&	 27.31(0.16)&  -10.27(0.01) &	4.99(0.04)  & 	5.14	& G&   & 	20.77(0.11)  &	-10.99(0.01) &	3.47(0.02) &	5.62	& G   \\
17143-3700	&	 11.09(0.26)&  -31.32(0.07) &	7.94(0.24)  & 	1.31	& B&   & 	 7.38(0.13)  &	-31.81(0.04) &	4.56(0.10) &	1.52	& G   \\
17158-3901	&	 35.28(0.66)&  -15.85(0.07) & 10.25(0.29)  & 	3.23	& R&   & 	20.34(0.14)  &	-16.19(0.00) &	6.08(0.05) &	3.14	& G   \\
17160-3707	&	 18.45(0.43)&  -69.65(0.06) &	7.47(0.19)  & 	2.32	& B&   & 	10.52(0.14)  &	-69.85(0.03) &	4.53(0.07) &	2.18	& G   \\
17175-3544	&  91.17(1.20)&   -5.74(0.06) & 11.00(0.25)  & 	7.79	& R&   & 	84.85(0.19)  &	 -6.54(0.01) &	6.55(0.02) &	12.17 & G   \\
17204-3636	&	 14.48(0.82)&  -17.76(0.10) &	5.66(0.45)  & 	2.40	& B&   & 	 6.05(0.16)  &	-17.94(0.05) &	3.66(0.12) &	1.55	& G   \\
17220-3609	&	 41.77(0.40)&  -95.04(0.06) & 12.73(0.23)  & 	3.08	& B&   & 	35.20(0.16)  &	-94.67(0.02) &	6.65(0.04) &	4.97	& G   \\
17233-3606	& 182.37(0.81)&   -4.15(0.04) & 21.04(0.15)  & 	8.14	& R&   & 113.92(0.84)  &	 -3.16(0.03) & 10.33(0.14) &	10.36 & R   \\
17244-3536	&	  5.39(0.12)&   -9.36(0.04) &	3.87(0.10)  & 	1.31	& G&   & 	 2.63(0.10)  &	-10.50(0.05) &	2.84(0.13) &	0.87	& G   \\
17258-3637	&	 46.49(1.51)&  -11.68(0.06) &	7.64(0.19)  & 	5.72	& R&   & 	35.66(0.29)  &	-12.11(0.02) &	5.38(0.04) &	6.22	& R   \\
17269-3312	&		3.36(0.22)&  -20.70(0.20) &	5.57(0.82)  & 	0.57	& G&   & 	 0.69(0.08)  &	-21.60(0.17) &	3.06(0.40) &	0.21	& G   \\
17271-3439	&	 30.21(2.53)&  -15.86(0.43) & 10.27(3.36)  & 	2.76	& B&   & 	24.24(0.33)  &	-15.40(0.03) &	4.55(0.07) &	5.01	& G   \\
17278-3541	&   7.40(0.23)&    0.73(0.07) &	4.71(0.21)  & 	1.48	& G&   & 	 3.19(0.14)  &	  0.35(0.08) &	3.58(0.15) &	0.84	& G   \\
17439-2845	&   4.54(0.15)&   18.76(0.07) &	4.32(0.17)  & 	0.99	& G&   & 	 1.56(0.13)  &	 18.15(0.11) &	3.90(0.55) &	0.38	& G   \\
17441-2822	& 186.66(40.2)&   51.49(1.53) & 15.12(1.81)  & 	11.60	& B&   & 262.61(1.23)  &	 60.05(0.03) & 22.95(0.10) &	10.75 & B   \\
17455-2800	&	 17.20(0.15)&  -14.19(0.02) &	5.62(0.06)  & 	2.88	& G&   & 	10.28(0.13)  &	-14.79(0.02) &	3.87(0.06) &	2.50	& G   \\
17545-2357	&   5.85(0.11)&    9.16(0.04) &	3.98(0.09)  & 	1.38	& G&   & 	 2.58(0.08)  &	  8.57(0.04) &	2.99(0.10) &	0.81	& G   \\
17589-2312	&   6.92(0.49)&   21.64(0.18) &	6.07(0.25)  & 	1.07	& R&   & 	 1.79(0.07)  &	 20.81(0.08) &	4.05(0.17) &	0.42	& G   \\
17599-2148	&  12.69(0.42)&   19.75(0.06) &	5.34(0.12)  & 	2.23	& R&   & 	 6.28(0.09)  &	 19.16(0.04) &	4.79(0.08) &	1.23	& G   \\
18032-2032	&  47.73(0.78)&    5.86(0.10) & 12.55(0.26)  & 	3.57	& R&   & 	32.22(0.46)  &	  4.49(0.05) &	8.00(0.17) &	3.78	& R   \\
18056-1952	&  51.66(1.12)&   66.70(0.10) & 11.03(0.33)  & 	4.40	& B&   & 	26.98(0.43)  &	 66.75(0.07) &	9.34(0.17) &	2.71	& B   \\
18075-2040	&    --        &    --        &   --        &    --  & --&   &     --         &      --      &    --      &  --  & --   \\
18079-1756	&  14.09(0.16)&   18.39(0.02) &	4.58(0.08)  & 	2.89	& G&   & 	 8.67(0.10)  &	 17.68(0.01) &	2.70(0.04) &	3.02	& G   \\
18089-1732	&  26.06(0.96)&   33.52(0.06) &	6.39(0.23)  & 	3.83	& B&   & 	16.48(0.14)  &	 33.13(0.02) &	4.65(0.06) &	3.33	& B   \\
18110-1854	&  18.81(0.29)&   39.04(0.03) &	5.90(0.14)  & 	2.99	& R&   & 	 8.88(0.08)  &	 38.55(0.02) &	3.72(0.04) &	2.24	& G   \\
18116-1646	&  23.90(0.53)&   50.25(0.03) &	6.71(0.12)  & 	3.35	& B&   & 	21.86(1.22)  &	 49.47(0.06) &	4.02(0.13) &	5.11	& B   \\
18117-1753	&  34.21(0.87)&   38.00(0.09) &	9.14(0.29)  & 	3.52	& R&   & 	20.85(0.28)  &	 36.46(0.04) &	6.47(0.06) &	3.03	& G   \\
18134-1942	&  10.81(0.12)&   10.74(0.02) &	3.50(0.05)  & 	2.90	& G&   & 	 5.15(0.07)  &	 10.14(0.02) &	2.44(0.04) &	1.99	& G   \\
18139-1842	&  13.05(0.84)&   39.98(0.05) &	4.44(0.21)  & 	2.76	& R&   & 	 7.44(0.12)  &	 39.61(0.03) &	3.39(0.06) &	2.06	& G   \\
18159-1648	&  39.77(0.77)&   23.88(0.07) &	7.66(0.26)  & 	4.88	& R&   & 	28.10(0.13)  &	 22.38(0.01) &	5.42(0.03) &	4.87	& G   \\
18182-1433	&  14.95(0.48)&   60.11(0.10) &	7.39(0.29)  & 	1.90	& B&   & 	 6.99(0.25)  &	 59.51(0.10) &	5.47(0.18) &	1.20	& G   \\
18223-1243	&   3.85(0.14)&   45.97(0.07) &	4.29(0.19)  & 	0.84	& G&   & 	 1.67(0.09)  &	 45.15(0.06) &	2.38(0.18) &	0.66	& G   \\
18228-1312	&  11.00(0.42)&   33.64(0.08) &	5.44(0.16)  & 	1.90	& B&   & 	 4.62(0.33)  &	 33.69(0.12) &	3.86(0.17) &	1.13	& R   \\
18236-1205	&  17.34(0.47)&   27.81(0.10) &	8.51(0.47)  & 	1.92	& R&   & 	 6.17(0.25)  &	 26.81(0.09) &	5.37(0.29) &	1.08	& G   \\
18264-1152	&  32.26(0.59)&   45.38(0.07) &	8.46(0.20)  & 	3.58	& G&   & 	12.06(0.21)  &	 43.95(0.03) &	4.49(0.10) &	2.52	& G   \\
18290-0924	&   --         &      --      &    --       &   --   & --&   &     --         &      --      &    --      &  --  & --   \\
18308-0503	&   2.89(0.14)&   43.56(0.07) &	3.09(0.19)  & 	0.88	& G&   & 	 0.71(0.07)  &	 43.01(0.08) &	1.59(0.20) &	0.42	& G   \\
18311-0809	&	 13.71(0.48)&  113.50(0.10) &	5.62(0.24)  & 	2.29	& G&   & 	 4.46(0.11)  &	112.99(0.04) &	3.81(0.11) &	1.10	& G   \\
18314-0720	&		3.75(0.17)&  102.14(0.17) &	7.81(0.44)  & 	0.45	& G&   & 	 1.38(0.12)  &	100.96(0.11) &	2.91(0.33) &	0.45	& G   \\
18316-0602	&  17.96(0.31)&   45.28(0.05) &	6.65(0.15)  & 	2.54	& G&   & 	13.20(0.16)  &	 42.80(0.03) &	6.19(0.10) &	2.00	& G   \\
18317-0513	&   8.52(0.14)&   42.21(0.03) &	3.67(0.07)  & 	2.18	& G&   & 	 4.26(0.09)  &	 41.58(0.03) &	2.45(0.06) &	1.63	& G   \\
18317-0757	&  10.86(0.20)&   80.52(0.05) &	5.32(0.12)  & 	1.92	& G&   & 	 7.00(0.08)  &	 80.07(0.02) &	3.09(0.04) &	2.13	& G   \\
18341-0727	&		8.93(0.25)&  113.49(0.09) &	6.19(0.27)  & 	1.36	& G&   & 	 2.30(0.11)  &	112.92(0.09) &	3.57(0.20) &	0.61	& G   \\
18411-0338	&		7.39(0.35)&  103.68(0.10) &	4.84(0.31)  & 	1.44	& G&   & 	 7.58(0.51)  &	103.66(0.12) &	5.02(0.46) &	1.42	& G   \\
18434-0242	&  28.42(0.17)&   98.06(0.01) &	5.08(0.04)  & 	5.25	& G&   & 	50.68(0.76)  &	 98.77(0.05) &	7.17(0.14) &	6.64	& G   \\
18440-0148	&   2.82(0.16)&   98.28(0.12) &	4.12(0.27)  & 	0.64	& G&   & 	   --         &      --      &    --      &  --   & --   \\
18445-0222	&   5.79(0.17)&   87.21(0.06) &	4.48(0.15)  & 	1.22	& G&   & 	 3.44(0.11)  &	 87.05(0.05) &	3.19(0.12) &	1.01	& G   \\
18461-0113	&  14.34(0.37)&   97.10(0.09) &	7.61(0.37)  & 	1.77	& R&   & 	 8.50(0.14)  &	 96.25(0.04) &	4.40(0.09) &	1.81	& G   \\
18469-0132	&  12.31(0.26)&   86.87(0.09) &	8.21(0.20)  & 	1.41	& G&   & 	 7.03(0.12)  &	 86.51(0.04) &	5.07(0.10) &	1.30	& G   \\
18479-0005	&  25.37(0.69)&   15.10(0.08) & 10.22(0.35)  & 	2.33	& R&   & 	17.67(0.20)  &	 14.62(0.05) &	8.16(0.10) &	2.03	& G   \\
18502+0051	&    --        &     --       &   --        &    --  & --&   &      --        &       --     &    --      &   -- & --   \\
18507+0110	&  74.45(0.82)&   58.46(0.03) &	8.30(0.09)  & 	8.43	& B&   & 	48.73(0.51)  &	 58.47(0.03) &	6.29(0.06) &	7.28	& B   \\
18507+0121	&  14.51(0.30)&   58.19(0.06) &	6.20(0.20)  & 	2.20	& R&   & 	 3.17(0.15)  &	 57.89(0.08) &	3.77(0.31) &	0.79	& G   \\
18517+0437	&  11.55(0.36)&   44.08(0.06) &	5.31(0.18)  & 	2.04	& G&   & 	 5.54(0.12)  &	 43.70(0.04) &	3.33(0.09) &	1.56	& G   \\
18530+0215	&   7.62(0.27)&   77.78(0.08) &	4.89(0.21)  & 	1.46	& G&   & 	 4.38(0.13)  &	 77.26(0.04) &	2.97(0.10) &	1.38	& G   \\
19078+0901	& 151.74(3.32)&    7.24(0.09) & 18.21(0.30)  & 	7.83	& B&   & 	97.37(0.77)  &	  6.59(0.04) & 13.35(0.10) &	6.85	& B   \\
19095+0930	&  23.02(0.59)&   44.70(0.09) &	8.09(0.18)  & 	2.67	& R&   & 	11.31(0.16)  &	 44.01(0.04) &	5.79(0.10) &	1.83	& G   \\
19097+0847	&   9.05(0.35)&   57.26(0.12) &	6.89(0.38)  & 	1.23	& G&   & 	 3.65(0.16)  &	 57.23(0.08) &	3.96(0.22) &	0.87	& G   \\
\end{longtable*}

\clearpage

\begin{longtable*}{ccccccccc}\centering
\tablecolumns{9} \tablewidth{0pc}
\tablecaption{Derived Parameters} \tablehead{
\colhead{IRAS}  & L$_{IR}$ &  L$_{TIR}$ & L$^{\prime}$(HCN)  &L$^{\prime}$(CS) &    $\sigma_{vir}$ & log(SFR)  & log($\Sigma_{dense}$) & log($\Sigma_{SFR}$)\\
  & \colhead{(L$_{\sun}$) } & \colhead{(L$_{\sun}$) }
  & \colhead{(K~km~s$^{-1}$~pc$^{2}$)} & \colhead{(K~km~s$^{-1}$~pc$^{2}$)} & \colhead{(km~s$^{-1}$)} & (M$_{\sun}$~yr$^{-1}$)  & (M$_{\sun}$~pc$^{-2}$) & (M$_{\sun}$~Myr$^{-1}$~pc$^{-2}$)
  } 	
08076-3556	 &  1.09E+01     & 1.60E+01   &                        &                & 0.47   & -8.63    & 3.14	& -0.23         \\
08303-4303	 &	 6.79E+03     & 6.70E+03   &    2.84(0.09)          &   1.75(0.05)   & 1.52   & -6.00   & 3.50	& 1.08  \\
08448-4343	 &	 6.63E+02     & 1.10E+03   &    0.73(0.01)          &   0.33(0.01)   & 0.87   & -6.79   & 3.35	& 0.97  \\
08470-4243	 &	 1.23E+04     & 1.10E+04   &    2.59(0.10)          &   1.35(0.05)   & 1.45   & -5.79   & 3.47	& 1.32  \\
09002-4732	 &	 4.61E+04     & 3.90E+04   &    3.03(0.03)          &   2.01(0.02)   & 1.77   & -5.24   & 3.76	& 2.11  \\
09018-4816	 &	 5.05E+04     & 5.20E+04   &   10.64(0.55)          &   7.63(0.09)   & 2.52   & -5.11   & 3.81	& 1.70  \\
09094-4803	 &  4.73E+04     & 4.00E+04   &                        &                & 1.70   & -5.23    & 2.96	& 0.59  \\
10365-5803	 &	 2.04E+04     & 1.90E+04   &    3.86(0.15)          &   1.70(0.19)   & 1.75   & -5.55   & 3.49	& 1.26  \\
11298-6155	 &	 1.91E+05     & 1.70E+05   &   24.65(0.96)          &  16.04(0.90)   & 2.39   & -4.60   & 3.27	& 1.24  \\
11332-6258	 &	 5.42E+03     & 5.30E+03   &    0.31(0.05)          &   0.29(0.03)   & 1.16   & -6.11   & 3.34	& 1.12  \\
11590-6452	 &	 3.85E+01     & 5.30E+01   &    0.01(0.00)          &                & 0.80   & -8.11   & 3.69	& 0.47  \\
12320-6122	 &	 2.64E+05     & 2.20E+05   &   25.87(0.69)          &  16.08(0.32)   & 2.84   & -4.49   & 3.68	& 1.87  \\
12326-6245	 &	 3.31E+05     & 2.70E+05   &   68.27(0.70)          &  56.13(0.30)   & 3.23   & -4.40   & 3.84	& 2.07  \\
12383-6128	 &	 4.95E+04     & 4.90E+04   &   19.43(0.55)          &  14.49(0.48)   & 1.90   & -5.14   & 3.10	& 0.76  \\
12572-6316	 &	 5.43E+03     & 5.60E+03   &    2.18(0.09)          &   1.22(0.07)   & 1.25   & -6.08   & 3.15	& 0.64  \\
13079-6218	 &  2.42E+05     & 2.80E+05   &   85.68(1.07)          &  54.36(0.58)   & 3.52   & -4.38    & 3.82	& 1.88  \\
13080-6229	 &	 1.95E+05     & 3.20E+05   &   14.96(0.27)          &   7.73(0.15)   & 1.92   & -4.32   & 3.45	& 2.26  \\
13111-6228	 &	 1.47E+05     & 1.30E+05   &   28.59(0.38)          &  23.08(0.22)   & 2.37   & -4.72   & 3.49	& 1.57  \\
13134-6242	 &	 3.51E+04     & 3.10E+04   &   11.01(0.36)          &   5.69(0.09)   & 1.97   & -5.34   & 3.66	& 1.61  \\
13140-6226	 &	 7.99E+03     & 7.80E+03   &    5.90(0.16)          &   2.71(0.13)   & 1.40   & -5.94   & 3.29	& 0.86  \\
13291-6229	 &	 5.23E+04     & 4.40E+04   &    7.35(0.35)          &   4.77(0.10)   & 1.20   & -5.19   & 2.96	& 1.23  \\
13291-6249	 &	 2.81E+04     & 2.50E+04   &    9.41(0.06)          &   5.46(0.06)   & 1.62   & -5.43   & 3.47	& 1.48  \\
13295-6152	 &	 6.50E+03     & 7.00E+03   &    3.23(0.29)          &                & 1.56   & -5.98   & 3.10	& 0.26  \\
13471-6120	 &	 3.12E+05     & 2.70E+05   &   23.50(0.30)          &  16.80(0.21)   & 2.70   & -4.40   & 3.62	& 1.92  \\
13484-6100	 &	 8.25E+04     & 7.10E+04   &   34.69(1.04)          &  23.53(0.49)   & 2.06   & -4.98   & 3.40	& 1.38  \\
14013-6105	 &	 2.04E+05     & 1.70E+05   &   61.46(1.41)          &  35.19(0.26)   & 2.54   & -4.60   & 3.53	& 1.66  \\
14050-6056	 &	              & 7.20E+04   &    9.61(0.18)          &   7.76(0.12)   & 1.83   & -4.97   & 3.35	& 1.49  \\
14164-6028	 &	 1.02E+04     & 9.30E+03   &    4.50(0.14)          &   2.74(0.06)   & 0.90   & -5.86   & 2.90	& 0.92  \\
14206-6151	 &	 1.29E+04     & 1.30E+04   &    1.08(0.13)          &   0.46(0.13)   & 1.46   & -5.72   & 3.15	& 0.75  \\
14212-6131	 &	 1.74E+04     & 1.20E+05   &   13.25(0.39)          &   4.47(0.29)   & 2.28   & -4.75   & 3.51	& 1.64  \\
14382-6017	 &	 8.55E+04     & 7.50E+04   &   15.68(0.81)          &  10.59(0.26)   & 1.62   & -4.95   & 3.11	& 1.23  \\
14453-5912	 &	 2.98E+04     & 2.80E+04   &    7.16(0.33)          &   3.88(0.08)   & 1.51   & -5.38   & 3.28	& 1.27  \\
14498-5856	 &	 3.87E+04     & 3.50E+04   &   18.71(0.31)          &  11.41(0.17)   & 2.11   & -5.29   & 3.55	& 1.32  \\
15122-5801	 &	 8.09E+04     & 7.20E+04   &    5.02(0.19)          &   3.78(0.17)   & 1.71   & -4.97   & 3.26	& 1.42  \\
15254-5621	 &	 1.70E+05     & 1.20E+05   &   24.66(0.28)          &  14.49(0.21)   & 2.74   & -4.75   & 3.66	& 1.63  \\
15290-5546	 &  4.56E+05     & 3.90E+05   &   78.59(1.76)          &  55.54(0.58)   & 3.32   & -4.24    & 3.71	& 1.90  \\
15384-5348	 &	 1.22E+05     & 1.00E+05   &    8.19(0.20)          &   4.81(0.13)   & 2.06   & -4.83   & 3.40	& 1.53  \\
15394-5358	 &	 1.50E+04     & 1.50E+04   &   20.68(0.45)          &   7.73(0.10)   & 3.39   & -5.65   & 4.06	& 1.15  \\
15408-5356	 &	 2.24E+05     & 1.90E+05   &   43.85(0.81)          &  28.67(0.98)   & 2.76   & -4.55   & 3.75	& 1.99  \\
15411-5352	 &	 1.57E+05     & 1.40E+05   &   13.39(0.24)          &   6.37(0.11)   & 2.19   & -4.68   & 3.66	& 2.09  \\
15437-5343	 &	 6.72E+04     & 6.10E+04   &    2.64(0.26)          &                & 1.70   & -5.04   & 3.22	& 1.27  \\
15439-5449	 &	 4.35E+04     & 4.40E+04   &    7.48(0.28)          &   3.75(0.07)   & 1.73   & -5.19   & 3.39	& 1.45  \\
15502-5302	 &	 8.90E+05     & 7.50E+05   &   46.91(0.48)          &  31.70(0.37)   & 3.52   & -3.95   & 3.84	& 2.36  \\
15520-5234	 &  2.23E+05     & 2.00E+05   &   52.64(0.32)          &  47.85(0.13)   & 2.73   & -4.53    & 3.90	& 2.33  \\
15522-5411	 &	 2.22E+04     & 2.10E+04   &    6.46(0.60)          &   5.66(0.27)   & 1.38   & -5.51   & 3.05	& 0.85  \\
15557-5215	 &  2.00E+04     & 2.30E+04   &   12.81(0.43)          &   5.30(0.30)   & 2.69   & -5.47    & 3.68	& 0.98  \\
15567-5236	 &	 9.79E+05     & 8.00E+05   &   85.65(0.93)          &  69.97(0.66)   & 4.50   & -3.93   & 3.95	& 2.16  \\
15570-5227	 &	 1.35E+05     & 1.60E+05   &   14.73(0.41)          &  11.85(0.26)   & 2.20   & -4.63   & 3.31	& 1.42  \\
15584-5247	 &	 3.93E+04     & 3.60E+04   &    6.02(0.41)          &   3.38(0.37)   & 1.79   & -5.27   & 3.16	& 0.85  \\
15596-5301	 &	 6.71E+04     & 6.40E+04   &   33.72(2.56)          &  22.98(0.38)   & 2.25   & -5.02   & 3.49	& 1.35  \\
16026-5035	 &	 1.64E+05     & 1.40E+05   &   13.37(0.79)          &   8.67(0.35)   & 1.77   & -4.68   & 3.07	& 1.27  \\
16037-5223	 &	 1.31E+05     & 1.10E+05   &   20.44(0.45)          &  19.67(0.34)   & 2.81   & -4.79   & 3.71	& 1.64  \\
16060-5146	 &	 7.48E+05     & 7.50E+05   &   48.01(2.20)          &  27.09(0.84)   & 5.21   & -3.95   & 4.22	& 2.44  \\
16065-5158	 &  3.03E+05     & 2.90E+05   &   48.91(1.06)          &  45.44(0.32)   & 3.78   & -4.37    & 4.07	& 2.27  \\
16071-5142	 &  7.80E+04     & 8.30E+04   &   79.40(2.81)          &  49.59(0.55)   & 3.68   & -4.91    & 3.89	& 1.41  \\
16076-5134	 &  2.26E+05     & 2.00E+05   &  116.84(1.13)          &  91.94(1.25)   & 2.79   & -4.53    & 3.56	& 1.61  \\
16119-5048	 &	 3.47E+04     & 3.30E+04   &   15.82(0.50)          &   8.86(0.18)   & 2.14   & -5.31   & 3.60	& 1.37  \\
16132-5039	 &	 7.20E+04     & 6.30E+04   &    3.91(0.15)          &   2.07(0.10)   & 1.65   & -5.03   & 3.26	& 1.43  \\
16158-5055	 &	 1.99E+05     & 1.60E+05   &   15.88(0.39)          &  11.39(0.24)   & 2.56   & -4.63   & 3.52	& 1.60  \\
16164-5046	 &	 4.51E+05     & 4.00E+05   &   56.74(1.18)          &  49.60(0.30)   & 4.35   & -4.23   & 4.24	& 2.51  \\
16172-5028	 &	 4.69E+05     & 4.40E+05   &   87.43(1.14)          & 112.92(1.17)   & 4.23   & -4.19   & 3.82	& 1.77  \\
16177-5018	 &	 3.83E+05     & 3.40E+05   &    4.79(0.44)          &   3.71(0.32)   & 4.66   & -4.30   & 4.26	& 2.36  \\
16272-4837	 &	 2.52E+04     & 3.00E+04   &   22.32(0.62)          &   9.11(0.15)   & 2.59   & -5.35   & 3.77	& 1.34  \\
16297-4757	 &	 5.77E+05     & 5.10E+05   &   87.74(2.00)          & 105.74(1.33)   & 3.28   & -4.12   & 3.43	& 1.49  \\
16304-4710	 &	 1.08E+05     & 1.00E+05   &   16.42(1.09)          &   6.89(1.00)   & 1.86   & -4.83   & 2.93	& 0.75  \\
16313-4729	 &	 1.09E+06     & 9.30E+05   &   73.41(2.00)          &  37.58(2.00)   & 3.41   & -3.86   & 3.35	& 1.52  \\
16318-4724	 &	 2.64E+05     & 2.50E+05   &   38.09(0.96)          &  28.61(0.42)   & 3.20   & -4.43   & 3.71	& 1.77  \\
16330-4725	 &	 2.27E+06     & 1.90E+06   &  297.54(2.59)          & 222.09(1.98)   & 3.01   & -3.55   & 3.19	& 1.73  \\
16344-4658	 &  5.28E+04     & 5.10E+04   &   21.15(0.50)          &  16.60(0.70)   & 2.26   & -5.12    & 3.49	& 1.25  \\
16348-4654	 &  5.45E+05     & 6.50E+05   &  264.00(4.19)          & 147.83(1.76)   & 5.58   & -4.02    & 3.92	& 1.65  \\
16351-4722	 &	 1.09E+05     & 9.30E+04   &   36.73(0.69)          &  33.25(0.40)   & 2.34   & -4.86   & 3.73	& 1.94  \\
16362-4639	 &	 2.01E+04     & 1.70E+04   &    3.19(0.15)          &   0.63(0.08)   & 0.96   & -5.60   & 2.75	& 0.79  \\
16372-4545	 &	 3.46E+04     & 3.70E+04   &   15.91(0.49)          &  13.04(0.39)   & 1.30   & -5.26   & 2.99	& 1.06  \\
16385-4619	 &  1.46E+05     & 1.30E+05   &   51.47(0.72)          &  46.73(0.39)   & 2.33   & -4.72    & 3.43	& 1.48  \\
16424-4531	 &	 1.92E+04     & 1.90E+04   &   13.10(0.23)          &   2.70(0.14)   & 1.82   & -5.55   & 3.42	& 1.06  \\
16445-4459	 &	 1.18E+05     & 1.10E+05   &   26.91(0.60)          &  11.81(0.39)   & 2.34   & -4.79   & 3.41	& 1.37  \\
16458-4512	 &	 4.99E+04     & 5.10E+04   &   36.72(0.76)          &   9.64(0.51)   & 2.54   & -5.12   & 3.47	& 1.00  \\
16484-4603	 &	 6.27E+04     & 5.50E+04   &   16.89(0.42)          &  10.99(0.16)   & 1.90   & -5.09   & 3.49	& 1.58  \\
16487-4423	 &	 4.79E+04     & 3.00E+04   &    8.89(0.22)          &   3.09(0.14)   & 1.64   & -5.35   & 3.20	& 1.01  \\
16489-4431	 &	 1.72E+04     & 1.50E+04   &    5.44(0.16)          &   2.41(0.10)   & 1.53   & -5.65   & 3.25	& 0.93  \\
16506-4512	 &	 1.17E+05     & 1.10E+05   &   22.55(0.44)          &  13.86(0.24)   & 2.34   & -4.79   & 3.52	& 1.59  \\
16524-4300	 &	 4.50E+04     & 4.20E+04   &   17.29(0.34)          &   8.93(0.10)   & 2.54   & -5.21   & 3.70	& 1.38  \\
16547-4247	 &  6.47E+04     & 6.30E+04   &   68.73(0.76)          &  31.36(0.22)   & 2.94   & -5.03    & 3.92	& 1.74  \\
16562-3959	 &	 5.63E+04     & 4.80E+04   &   16.66(0.11)          &   8.94(0.05)   & 3.71   & -5.15   & 4.29	& 1.97  \\
16571-4029	 &	 8.81E+04     & 7.60E+04   &   17.01(0.92)          &  11.65(0.22)   & 1.51   & -4.95   & 3.21	& 1.56  \\
17006-4215	 &	 4.58E+04     & 4.10E+04   &    6.33(0.10)          &   5.33(0.05)   & 2.29   & -5.22   & 3.81	& 1.76  \\
17008-4040	 &	 6.47E+04     & 5.20E+04   &   10.96(0.23)          &   6.85(0.08)   & 3.42   & -5.11   & 4.15	& 1.86  \\
17016-4124	 &  6.73E+04     & 5.90E+04   &   36.08(0.62)          &  14.23(0.18)   & 3.53   & -5.06    & 4.15	& 1.86  \\
17136-3617	 &	 1.17E+05     & 1.10E+05   &    8.11(0.05)          &   6.17(0.03)   & 2.46   & -4.79   & 3.90	& 2.25  \\
17143-3700	 &	 4.72E+04     & 4.10E+04   &   10.43(0.24)          &   6.94(0.12)   & 1.37   & -5.22   & 3.17	& 1.39  \\
17158-3901	 &  2.11E+04     & 2.20E+04   &    9.05(0.17)          &   5.22(0.04)   & 2.62   & -5.49    & 4.01	& 1.66  \\
17160-3707	 &	 4.47E+05     & 4.30E+05   &  122.66(2.86)          &  69.94(0.93)   & 4.59   & -4.20   & 3.72	& 1.41  \\
17175-3544	 &  7.56E+04     & 7.80E+04   &   12.06(0.16)          &  11.23(0.03)   & 3.32   & -4.94    & 4.37	& 2.52  \\
17204-3636	 &	 1.51E+04     & 1.50E+04   &   10.29(0.58)          &   4.30(0.11)   & 1.86   & -5.65   & 3.47	& 1.01  \\
17220-3609	 &  5.45E+05     & 6.80E+05   &  304.01(2.91)          & 256.19(1.16)   & 6.46   & -4.00    & 4.05	& 1.67  \\
17233-3606	 &  1.59E+04     & 1.70E+04   &    6.57(0.03)          &   4.10(0.03)   & 2.35   & -5.60    & 4.36	& 2.43  \\
17244-3536	 &	 2.11E+04     & 1.90E+04   &    2.55(0.06)          &   1.25(0.05)   & 1.37   & -5.55   & 3.28	& 1.29  \\
17258-3637	 &	 2.12E+05     & 1.90E+05   &   17.92(0.58)          &  13.75(0.11)   & 3.28   & -4.55   & 4.11	& 2.41  \\
17269-3312	 &	 5.52E+04     & 4.90E+04   &    4.50(0.29)          &   0.92(0.11)   & 1.68   & -5.14   & 3.24	& 1.26  \\
17271-3439	 &  2.78E+05     & 2.60E+05   &   38.82(3.25)          &  31.15(0.42)   & 5.32   & -4.42    & 4.24	& 1.98  \\
17278-3541	 &	 6.07E+03     & 5.80E+03   &    2.29(0.07)          &   0.99(0.04)   & 1.45   & -6.07   & 3.38	& 0.87  \\
17439-2845	 &	 5.14E+05     & 4.80E+05   &   33.53(1.11)          &  11.52(0.96)   & 3.13   & -4.15   & 3.39	& 1.45  \\
17441-2822	 &  3.70E+06     & 3.70E+06   & 2489.25(536.10)        &3501.96(16.40)  & 16.89  & -3.26    & 4.70	& 2.04   \\
17455-2800	 &	 8.67E+05     & 8.10E+05   &  191.52(1.67)          & 114.47(1.45)   & 4.13   & -3.92   & 3.54	& 1.51  \\
17545-2357	 &	 1.55E+04     & 1.40E+04   &    4.52(0.09)          &   2.00(0.06)   & 1.54   & -5.68   & 3.26	& 0.89  \\
17589-2312	 &	 2.34E+04     & 2.10E+04   &    5.88(0.42)          &   1.52(0.06)   & 1.06   & -5.51   & 2.97	& 1.14  \\
17599-2148	 &	 1.18E+05     & 1.10E+05   &   14.09(0.47)          &   6.98(0.10)   & 2.84   & -4.79   & 3.72	& 1.66  \\
18032-2032	 &  3.72E+05     & 3.20E+05   &   91.19(1.49)          &  61.56(0.88)   & 3.28   & -4.32    & 3.77	& 1.97  \\
18056-1952	 &  3.58E+05     & 3.50E+05   &  113.09(2.45)          &  59.06(0.94)   & 3.67   & -4.29    & 3.82	& 1.91  \\
18075-2040	 &  2.46E+04     & 2.40E+04   &                        &                & 1.44   & -5.45    & 2.99	& 0.70  \\
18079-1756	 &	 2.42E+04     & 2.00E+04   &    7.51(0.09)          &   4.62(0.05)   & 0.66   & -5.53   & 2.61	& 1.24  \\
18089-1732	 &	 5.13E+04     & 5.20E+04   &   21.91(0.81)          &  13.86(0.12)   & 3.01   & -5.11   & 3.87	& 1.53  \\
18110-1854	 &	 1.09E+05     & 9.60E+04   &   19.42(0.30)          &   9.17(0.08)   & 2.30   & -4.85   & 3.60	& 1.71  \\
18116-1646	 &	 2.06E+05     & 1.80E+05   &   35.09(0.78)          &  32.10(1.79)   & 2.84   & -4.57   & 3.68	& 1.78  \\
18117-1753	 &	 1.11E+05     & 1.10E+05   &   37.47(0.95)          &  22.83(0.31)   & 3.13   & -4.79   & 3.83	& 1.70  \\
18134-1942	 &	 6.56E+03     & 6.80E+03   &    3.87(0.04)          &   1.84(0.03)   & 1.45   & -6.00   & 3.37	& 0.91  \\
18139-1842	 &	 6.62E+04     & 5.80E+04   &   14.68(0.94)          &   8.37(0.13)   & 1.38   & -5.07   & 3.12	& 1.43  \\
18159-1648	 &	 2.18E+04     & 2.50E+04   &   16.51(0.32)          &  11.67(0.05)   & 3.14   & -5.43   & 4.06	& 1.52  \\
18182-1433	 &	 2.69E+04     & 2.80E+04   &   16.30(0.52)          &   7.62(0.27)   & 1.90   & -5.38   & 3.43	& 1.18  \\
18223-1243	 &	 2.50E+04     & 2.70E+04   &    5.06(0.18)          &   2.20(0.12)   & 1.95   & -5.40   & 3.36	& 0.97  \\
18228-1312	 &	 5.84E+04     & 5.90E+04   &    9.16(0.35)          &   3.85(0.27)   & 1.91   & -5.06   & 3.44	& 1.52  \\
18236-1205	 &	 6.92E+03     & 8.20E+03   &   10.81(0.29)          &   3.85(0.16)   & 2.10   & -5.92   & 3.58	& 0.78  \\
18264-1152	 &	 1.38E+04     & 1.60E+04   &   23.24(0.42)          &   8.69(0.15)   & 2.53   & -5.63   & 3.76	& 1.09  \\
18290-0924	 &  3.23E+04     & 3.40E+04   &                        &                & 1.97   & -5.30    & 3.26	& 0.84  \\
18308-0503	 &	 2.30E+04     & 2.10E+04   &    2.47(0.12)          &   0.61(0.06)   & 1.40   & -5.51   & 3.17	& 1.06  \\
18311-0809	 &	 3.90E+05     & 3.60E+05   &  103.25(3.62)          &  33.59(0.83)   & 3.30   & -4.27   & 3.44	& 1.35  \\
18314-0720	 &	 5.01E+05     & 4.40E+05   &   30.39(1.38)          &  11.18(0.97)   & 2.77   & -4.19   & 3.24	& 1.33  \\
18316-0602	 &	 3.03E+04     & 2.90E+04   &   12.56(0.22)          &   9.23(0.11)   & 2.37   & -5.37   & 3.69	& 1.32  \\
18317-0513	 &	 2.96E+04     & 2.70E+04   &    7.17(0.12)          &   3.59(0.08)   & 1.96   & -5.40   & 3.47	& 1.19  \\
18317-0757	 &	 1.63E+05     & 1.40E+05   &   23.20(0.43)          &  14.95(0.17)   & 2.29   & -4.68   & 3.40	& 1.49  \\
18341-0727	 &	 1.91E+05     & 1.90E+05   &   42.21(1.18)          &  10.87(0.52)   & 3.44   & -4.55   & 3.57	& 1.24  \\
18411-0338	 &	 1.84E+05     & 1.60E+05   &   20.48(0.97)          &  21.00(1.41)   & 2.32   & -4.63   & 3.38	& 1.47  \\
18434-0242	 &	 1.31E+06     & 1.10E+06   &   80.12(0.48)          & 142.87(2.14)   & 3.76   & -3.79   & 3.83	& 2.38  \\
18440-0148	 &	 8.76E+04     & 7.90E+04   &   12.46(0.71)          &                & 1.03   & -4.93   & 2.58	& 0.98  \\
18445-0222	 &	 7.56E+04     & 7.00E+04   &   13.92(0.41)          &   8.27(0.26)   & 1.80   & -4.98   & 3.17	& 1.14  \\
18461-0113	 &	 6.35E+04     & 6.30E+04   &   31.38(0.81)          &  18.60(0.31)   & 2.24   & -5.03   & 3.41	& 1.21  \\
18469-0132	 &	 1.58E+05     & 1.40E+05   &   28.24(0.60)          &  16.13(0.28)   & 2.07   & -4.68   & 3.31	& 1.48  \\
18479-0005	 &  1.61E+06     & 1.40E+06   &  230.81(6.28)          & 160.76(1.82)   & 5.03   & -3.68    & 3.82	& 1.96  \\
18502+0051	 &  2.22E+05     & 2.00E+05   &                        &                & 4.17   & -4.53    & 3.79	& 1.39  \\
18507+0110	 &	 5.35E+05     & 7.50E+05   &  179.35(1.98)          & 117.39(1.23)   & 4.13   & -3.95   & 3.85	& 2.10  \\
18507+0121	 &	 3.13E+04     & 3.40E+04   &   23.74(0.49)          &   5.19(0.25)   & 2.61   & -5.30   & 3.55	& 0.95  \\
18517+0437	 &	 1.80E+04     & 2.20E+04   &    7.26(0.23)          &   3.48(0.08)   & 1.81   & -5.49   & 3.48	& 1.25  \\
18530+0215	 &	 9.72E+04     & 9.00E+04   &   15.03(0.53)          &   8.64(0.26)   & 1.99   & -4.88   & 3.31	& 1.35  \\
19078+0901	 &  4.89E+06     & 5.60E+06   & 1474.94(32.2)          & 946.45(7.48)   & 15.09  & -3.08    & 4.73	& 2.47  \\
19095+0930	 &	 4.83E+04     & 4.20E+04   &   14.11(0.36)          &   6.93(0.10)   & 1.79   & -5.21   & 3.50	& 1.60  \\
19097+0847	 &	              & 4.40E+04   &   13.76(0.53)          &   5.55(0.24)   & 2.36   & -5.19   & 3.51	& 1.14  \\
\end{longtable*}

\begin{deluxetable*}{ccccccccc}
\tabletypesize{\scriptsize} \tablecolumns{10} \tablewidth{0pc}
\tablecaption{Parameters of L$_{TIR}$--L$\arcmin_{mol}$ correlations for each sub-sample } \tablehead{
\colhead{Group}  & & & CS (7-6) & & & &  HCN (4-3)& \\
\cline{3-5}\cline{7-9}\\
 & & \colhead{$\alpha$} & \colhead{$\beta$}  & \colhead{R} &  & \colhead{$\alpha$} & \colhead{$\beta$}  & \colhead{R}   } \startdata
$<$n$>$=$3.7\times10^{4}$ cm$^{-3}$          &       &  0.89(0.06)   &   4.03(0.08)   &   0.92  &       &  0.97(0.06)   &  3.72(0.09)   &  0.92  \\
$<$n$>$=$2.8\times10^{5}$ cm$^{-3}$          &       &  0.80(0.07)   &   3.98(0.10)   &   0.88  &       &  0.97(0.09)   &  3.63(0.14)   &  0.83  \\
$<$T$_{d}>$=29 K                             &       &  0.97(0.05)   &   3.67(0.07)   &   0.94  &       &  0.90(0.06)   &  3.52(0.08)   &  0.92  \\
$<$T$_{d}>$=36 K                             &       &  0.74(0.06)   &   4.23(0.09)   &   0.87  &       &  0.77(0.08)   &  4.12(0.12)   &  0.82  \\
$<$L/M$>$=29 L$_{\sun}$/M$_{\sun}$           &       &  0.88(0.04)   &   3.74(0.06)   &   0.96  &       &  1.00(0.05)   &  3.34(0.07)   &  0.95  \\
$<$L/M$>$=107 L$_{\sun}$/M$_{\sun}$          &       &  0.77(0.07)   &   4.27(0.10)   &   0.84  &       &  0.77(0.07)   &  4.27(0.10)   &  0.84  \\
$<\sigma_{CS}/\sigma_{vir}>$=0.55            &       &  0.71(0.05)   &   4.30(0.08)   &   0.91  &       &     &     &    \\
$<\sigma_{CS}/\sigma_{vir}>$=1.14            &       &  0.92(0.06)   &   3.75(0.08)   &   0.91  &       &     &     &    \\
$<\sigma_{HCN}/\sigma_{vir}>$=0.86           &       &     &      &     &       &  0.81(0.07)   &  4.00(0.11)   &  0.88  \\
$<\sigma_{HCN}/\sigma_{vir}>$=1.65           &       &     &      &     &       &  0.93(0.07)   &  3.55(0.10)   &  0.89  \\
G                                            &       &  0.89(0.05)   &   3.95(0.06)   &   0.87  &       &  0.99(0.06)   &  3.77(0.07)   &  0.90  \\
B+R                                          &       &  0.77(0.09)   &   4.08(0.17)   &   0.88  &       &  0.98(0.06)   &  3.52(0.10)   &  0.89
\enddata
\end{deluxetable*}

\begin{deluxetable*}{cccccccccccc}
\tabletypesize{\scriptsize} \tablecolumns{13} \tablewidth{0pc}
\tablecaption{Parameters of correlations for different transitions} \tablehead{
\colhead{Group} &E$_{u}$ & n$_{crit}$\tablenotemark{a} & n$_{eff}$\tablenotemark{b} & & & L$_{TIR}$--L$\arcmin_{mol}$\tablenotemark{c} & & & &  M$_{clump}$--L$\arcmin_{mol}$& \\
\cline{6-8}\cline{10-12}\\
 & \colhead{(K)}  & \colhead{(cm$^{-3}$)} & \colhead{(cm$^{-3}$)} & & \colhead{$\alpha$} & \colhead{$\beta$}  & \colhead{R} &  & \colhead{$\alpha$} & \colhead{$\beta$}  & \colhead{R}   } \startdata
CS (2-1)          & 7.1   & 1.0E+5 &1.2E+4 &     &  1.03(0.05)   &   3.25(0.11)   &   0.80  &       &  0.92(0.08)   &  1.25(0.16)   &  0.91  \\
CS (5-4)          & 35.3  & 1.7E+6 &2.5E+5 &     &  1.05(0.05)   &   3.77(0.08)   &   0.86  &       &  1.00(0.06)   &  1.74(0.08)   &  0.95  \\
CS (7-6)          & 65.8  & 4.4E+6 &2.1E+6 &     &  0.72(0.05)   &   4.23(0.07)   &   0.85  &       &  0.77(0.04)   &  2.29(0.05)   &  0.87  \\
HCN (1-0)         & 4.25  & 3.0E+5 &4.5E+3 &     &  1.07(0.06)   &   2.98(0.14)   &   0.85  &       &  0.99(0.07)   &  0.89(0.15)   &  0.94  \\
HCN (3-2)         & 25.52 & 1.0E+7 &7.3E+4 &     &  0.88(0.06)   &   3.94(0.11)   &   0.92  &       &  0.93(0.06)   &  1.58(0.09)   &  0.96  \\
HCN (4-3)         & 42.53 & 2.3E+7 &3.2E+5 &     &  0.74(0.06)   &   4.09(0.09)   &   0.79  &       &  0.85(0.04)   &  2.02(0.05)   &  0.89
\enddata
\tablenotetext{a}{critical density at 20 K, from \cite{shir15}}
\tablenotetext{b}{effective excitation density at 20 K, from \cite{shir15}}
\tablenotetext{c}{Parameters of L$_{TIR}$--L$\arcmin_{mol}$ for CS (2-1), (5-4) and HCN (1-0) and (3-2) are from \cite{wu10}. In \cite{wu10}, they only considered clumps with L$_{TIR}>$10$^{4.5}$ L$_{\sun}$. To compare with \cite{wu10}, we also only fitted L$_{TIR}$--L$\arcmin_{CS}$ of HCN (4-3) and CS (7-6) for clumps with L$_{TIR}>$10$^{4.5}$ L$_{\sun}$. }
\end{deluxetable*}


\begin{thebibliography}

\bibitem[Clements, Dunne, \& Eales (2010)]{clem10}Clements, D. L., Dunne, L., \& Eales, S., 2010, \mnras, 403, 274	

\bibitem[Clemens et al. (2013)]{clem13}Clemens, M. S., Negrello, M., De Zotti, G., et al. 2013, \mnras, 433, 695

\bibitem[Csengeri et al. (2016)]{csen16}Csengeri, T., Weiss, A., Wyrowski, F., et al. 2016, A\&A, 585, 104

\bibitem[da Cunha et al. (2010)]{da10}da Cunha, E., Eminian, C., Charlot, S., et al.,  2010, \mnras, 403, 1894

\bibitem[Daddi et al.\ (2010)]{daddi10}Daddi, E., Elbaz, D., Walter, F., et al. 2010, \apj, 714, L118

\bibitem[Dunne et al. (2000)]{dunne00}Dunne, L., Eales, S., Edmunds, Mi., et al. 2000, \mnras, 315, 115

\bibitem[Fa\'{u}ndez et al.\ (2004)]{fa04}Fa\'{u}ndez, S., Bronfman, L., Garay, G., et al.\ 2004, \aap, 426, 97

\bibitem[Gao \& Solomon\ (2004)]{gao04}Gao, Y., \& Solomon, P. M. 2004, ApJ, 606, 271

\bibitem[Genzel et al.\ (2010)]{gen10}Genzel, R., Tacconi, L. J., Gracia-Carpio, J., et al. 2010, \mnras, 407, 2091

\bibitem[Graci\'{a}-Carpio et al.\ (2008)]{gra08}Graci\'{a}-Carpio, J., Garc\'{\i}a-Burillo, S., Planesas, P., et al., 2008, \aap

\bibitem[Greve et al.\ (2014)]{gre14}Greve, T. R., Leonidaki, I., Xilouris, E. M., et al. 2014, \apj, 794, 142

\bibitem[Guilloteau \& Lucas(2000)]{gui00}Guilloteau, S. \& Lucas, R., 2000, in Astronomical Society of the Pacific Conference Series, Vol. 217, Imaging at Radio through Submillimeter
Wavelengths, ed. J. G. Mangum \& S. J. E. Radford, 299

\bibitem[Hjorth et al. (2014)]{hjor14}Hjorth, J., Gall, C., Micha{\l}owski, Micha{\l} J., 2014, \apj, 782L, 23

\bibitem[Juneau et al.\ (2009)]{jun09}Juneau, S., Narayanan, D. T., Moustakas, J., et al., 2009, \apj, 707, 1217

\bibitem[Kennicutt et al.\ (1998)]{ken98}Kennicutt, R. C., Jr. 1998, \apj, 498, 541

\bibitem[Kennicutt \& Evans (2012)]{ken12}Kennicutt, R. C. \& Evans, N. J., 2012, \araa, 50, 531

\bibitem[Lada, Lombardi, \& Alves\ (2010)]{lada10}Lada, C. J., Lombardi, M., \& Alves, J. F. 2010, ApJ, 724, 687

\bibitem[Lada et al.\ (2012)]{lada12}Lada, C. J., Forbrich, J., Lombardi, M., et al. 2012, \apj, 745, 190

\bibitem[Lada et al.\ (2013)]{lada13}Lada, C. J., Lombardi, M., Roman-Zuniga, C., et al., 2013, \apj, 778, 133

\bibitem[Liu, Wu \& Zhang (2013)]{liu13}Liu, T., Wu, Y., \& Zhang, H., 2013, \apj, 775, L2

\bibitem[Liu et al.\ (2015)]{liu15}Liu, D., Gao, Y., Isaak, K., et al. 2015, \apj, 810L, 14L

\bibitem[Ma, Tan \& Barnes (2013)]{ma13}Ma, B., Tan, J. C. \& Barnes, P. J., 2013, \apj, 779, 79

\bibitem[Magdis et al. (2013)]{mag13}Magdis, G. E., Rigopoulou, D., Helou, G., et al. 2013, \aap, 558, 136

\bibitem[Molinari et al. (2016)]{moli16}Molinari, S., Merello, M., Elia, D., et al. 2016, ApJL, in press, 2016arXiv160406192M

\bibitem[Mueller et al.\ (2002)]{mue02}Mueller, K. E., Shirley, Y. L., Evans, N. J., II, et al. 2002, \apjs,
143, 469

\bibitem[Narayanan et al.\ (2008)]{nara08}Narayanan, D., Cox, T. J., Shirley, Y., et al. 2008, \apj, 684, 996

\bibitem[Ossenkopf et al. (2010)]{Oss10}Ossenkopf, V., R\"{}llig, M., Simon, R., et al. 2010, \aap, 518, L79

\bibitem[Reiter et al. (2011)]{reiter11}Reiter, M., Shirley, Y. L., Wu, J., et al. 2011, \apjs, 195, 1

\bibitem[Rowlands et al. (2012)]{row12}Rowlands, K., Dunne, L., Maddox, S., et al. 2012, \mnras, 419, 2545

\bibitem[Santini et al. (2014)]{sant14}Santini, P., Maiolino, R., Magnelli, B., et al. 2014, \aap, 562, A30

\bibitem[Schmidt\ (1959)]{sch59}Schmidt, M. 1959, \apj, 129, 243

\bibitem[Scoville et al. (2016)]{scov16}Scoville, N., Sheth, K., Aussel, H., et al. 2016, \apj, 820, 83

\bibitem[Shirley\ (2015)]{shir15}Shirley, Yancy L., 2015, PASP, 127, 299

\bibitem[Stephens et al. (2016)]{step16}Stephens, I. W., Jackson, J. M., Whitaker, J. S., 2016, accepted to ApJ, 2016arXiv160309341S

\bibitem[van Kempen et al. (2009)]{van09}van Kempen, T. A., van Dishoeck, E. F., G\"{u}sten, R., et al. 2009, \aap, 501, 633

\bibitem[Vutisalchavakul \& Evans (2013)]{Vuti13}Vutisalchavakul, N., \& Evans, N. J., II, 2013, \apj, 765, 129

\bibitem[Wienen et al. (2015)]{wien15}Wienen, M., Wyrowski, F., Menten, K. M., et al. 2015, \aap, 579, 91

\bibitem[Williams, de Geus,\& Blitz (1994)]{wil94}Williams, J. P., de Geus, E. J., \& Blitz, L. 1994, \apj, 428, 693

\bibitem[Wu et al.\ (2005)]{wu05}Wu, J., Evans, N. J., II, Gao, Y., et al. 2005, \apj, 635, L173

\bibitem[Wu et al.\ (2010)]{wu10}Wu, J., Evans, N. J., II, Shirley, Y. L., et al. 2010, \apjs, 188, 313

\bibitem[Zhang et al.\ (2014)]{zhang14}Zhang, Z.-Y., Gao, Y., Henkel, C., et al. 2014, \apjl, 784, L31

\bibitem[Zhou et al.(1993)]{zhou93}Zhou, S., Evans, N. J., II, Koempe, C., \& Walmsley, C. M. 1993, \apj, 404, 232


\end{thebibliography}
\end{document}